\documentclass[manuscript]{emulateapj}
\usepackage{graphicx}
\slugcomment{Accepted by ApJ, to appear in v640, 20 March 2006}

\shorttitle{GDDS VI - Massive H$\delta$-Strong galaxies at $z \simeq 1$}
\shortauthors{D. Le Borgne et al.}

\newcommand{\Hdelta}{H$\delta$}
\newcommand{\HdeltaA}{H$\delta_\mathrm{A}$}
\newcommand{\HdeltaAmin}{H$\delta_\mathrm{A\,min}$}
\newcommand{\HdeltaEW}{EW(H$\delta$)}
\newcommand{\HDS}{H$\delta$-strong}
\newcommand{\OIIEW}{EW[\ion{O}{2}]}
\newcommand{\Dn}{D$_n4000$}
\newcommand{\ebv}{E(B-V)}
\newcommand{\Msel}{$M_\star>10^{10.2}$~M$_\odot$}
\newcommand{\fhds}{$f_\mathrm{HDS}$}
\begin{document}
\newcommand{\dd}{\mathrm{d}}
  
\title{Gemini Deep Deep Survey. VI. Massive H$\delta$-Strong galaxies at $z \simeq 1$}

\author{Damien Le Borgne\altaffilmark{1}, Roberto Abraham\altaffilmark{1}, Kathryne Daniel\altaffilmark{2}, Patrick J. McCarthy\altaffilmark{3}, Karl
  Glazebrook\altaffilmark{2}, Sandra Savaglio\altaffilmark{2}, David Crampton\altaffilmark{4}, 
  St\'ephanie Juneau\altaffilmark{5,4}, Ray G. Carlberg\altaffilmark{1}, Hsiao-Wen Chen\altaffilmark{6,9}, Ronald O. Marzke\altaffilmark{7},
  Kathy Roth\altaffilmark{8}, Inger J{\o}rgensen\altaffilmark{8}, Richard Murowinski\altaffilmark{4}}

\altaffiltext{1}{Department of Astronomy \& Astrophysics, University of Toronto,
        Toronto ON, M5S~3H8 Canada, [leborgne; abraham; carlberg]@astro.utoronto.ca}

\altaffiltext{2}{Department of Physics \& Astronomy, Johns Hopkins
University, Baltimore, MD 21218, [kdaniel; kgb; savaglio]@pha.jhu.edu}

\altaffiltext{3}{Carnegie Observatories, 813 Santa Barbara St,
        Pasadena, CA 91101, pmc2@ociw.edu}

\altaffiltext{4}{NRC Herzberg Institute for Astrophysics, 5071 W. Saanich Rd.,
Victoria, BC, Canada, [david.crampton; murowinski]@nrc-cnrc.gc.ca}

\altaffiltext{5}{D\'{e}partement de physique, Universit\'{e} de Montr\'{e}al, 2900,
 Bld. \'{E}douard-Montpetit, Montr\'{e}al, QC, Canada H3T 1J4, sjuneau@astro.umontreal.ca}

\altaffiltext{6}{Center for Space Sciences, Massachusetts Institute of
Technology, 70 Vassar St., Bld. 37, Cambridge, MA 02139, hchen@space.mit.edu}

\altaffiltext{7}{Department of Physics and Astronomy, San Francisco State University, San Francisco, CA 94132, marzke@stars.sfsu.edu}

\altaffiltext{8}{Gemini Observatory, 670 North A'ohoku Place, Hilo, HI 97620, [jorgensen; kroth]@gemini.edu}

\altaffiltext{9}{Hubble Fellow}

\begin{abstract}
We show that there has been a dramatic decline in the abundance of
massive galaxies with strong \Hdelta{} stellar absorption lines from
$z\sim 1.2$ to the present.  These ``\HDS{}'', or HDS, galaxies have
undergone a recent and rapid break in their star-formation
activity. Combining data from the Gemini Deep Deep and the Sloan
Digital Sky Surveys to make mass-matched samples (\Msel{}, with 25 and
$50,255$ galaxies, respectively), we find that the fraction of
galaxies in an HDS phase has decreased from about 50\% at $z=1.2$ to a
few percent today.  This decrease in fraction is due to an actual
decrease in the number density of massive HDS systems by a factor of
2-4, coupled with an increase in the number density of massive
galaxies by $\sim 30$ percent. We show that this result depends only
weakly on the threshold chosen for the \Hdelta{} equivalent width to
define HDS systems (if greater than 4\AA{}) and corresponds to a $(1+z)^{2.5\pm0.7}$ evolution.
Spectral synthesis studies of the high-redshift population using the
{\sc P\'egase} code, treating \HdeltaA{}, \OIIEW{}, \Dn{}, and
rest-frame colors, favor models in which the Balmer absorption
features in massive H$\delta$-strong systems  are the echoes of
intense episodes of star-formation that faded $\simeq 1$~Gyr prior to
the epoch of observation.  The $z=1.4-2$ epoch appears to correspond
to a time at which massive galaxies are in transition from a mode of
sustained star formation to a relatively quiescent mode with weak and
rare star-formation episodes. We argue that the most likely local
descendants of the distant massive HDS galaxies are passively evolving
massive galaxies in the field and small groups.
\end{abstract}


%%%%%%%%%%%%%%%%%%%%%%%%%%%%%%%%%%%%%%%%%%%%%%%%%%%%%%%%%%%%%%%%%%%%%%%%%%%%%%%%%
\section{Introduction}

The history of cosmic star-formation can be conveniently
encapsulated by the evolution of the star-formation rate density
(SFRD) as a function of redshift $z$. The SFRD is a fundamental quantity
that places stringent limits on models for galaxy formation
\citep[e.g.][]{Madauetal1998}. In hierarchical models, individual galaxies are the ephemeral
building blocks of more massive systems, and galaxy formation is a
continuous process.  In this picture, it is difficult to make an evolutionary connection between a
massive galaxy seen at low redshift and a very similar galaxy seen at
a substantially higher redshift, because today's massive galaxies were
in several pieces at high redshifts.  This difficulty in `connecting
the dots' between similar galaxies over a range of redshifts is
greatly reduced if massive galaxies form earlier and are longer-lived
than predicted by early generations of hierarchical models, as now appears to be the
case \citep[][hereafter Paper~III]{GDDS_PIII}.  In this view, a
detailed understanding of the formation and evolution of \emph{individual} massive
galaxies is entirely complementary to the study of volume-averaged quantities
such as the SFRD \citep[e.g.][hereafter Paper V]{GDDS_PV}.

% why we're interested in spectral indices
Broad-band colors and spectral features have long been used to probe
the star-formation history of individual galaxies, with mixed
success. Emission lines and blue starlight probe the current
star-formation rate. Constraints on the past history of star-formation
from the colors and spectral features of older stellar populations,
however, are hampered by the well-known age-metallicity degeneracy.
Fortunately, characterizing the {\em recent} ($\lesssim 2$~Gyr)
star-formation history of galaxies is more tractable.  In particular,
the \HdeltaA{} Lick index \citep{Worthey&Ottaviani1997} and the
4000-\AA{} break \citep{Dressler&Shectman1987}, when combined
together, have proved to be powerful diagnostics for probing recent
star-formation.  Such subtle diagnostics of recent star-formation can
only be probed by spectroscopy as they only have small effects on
broad-band colors. Galaxies which have experienced a break in
their star-formation history are particularly well identified by
such features. For instance, the evolution of the \HdeltaA{} Lick
index for a stellar population reaches
high values ($>5$~\AA, corresponding to a strong absorption line) only
when this population is dominated by A stars, which happens
exclusively for intermediate ages ($\sim 300$~Myr to $\sim 2$~Gyr)
under the condition that no significant star formation has occurred in
the last few hundred Myrs.  This behavior is predicted with good
confidence by the recent generation of galaxy evolution models, such
as {\sc P\'egase.2} \citep{FRV97,PEGASE2} and {\sc P\'egase-HR}
\citep{PEGASEHR} at low and high spectral resolutions,
respectively. The systems identified by this particularly deep
\Hdelta{} line are usually called ``\HDS{}''.

The \Hdelta{} line and the 4000-\AA{} break have long been used to probe
stellar populations, originally mostly in clusters of galaxies
\citep{Dressler&Gunn1983, Couch&Sharples1987,
Abrahametal1996, Poggiantietal1999}, and more recently also in field
galaxies \citep{Zabludoffetal1996, Zabludoff1999, Baloghetal1999,
Gotoetal2003, Quinteroetal2004, Tranetal2004, Yangetal2004}. In a
particularly significant recent development, \citet{Kauffmannetal2003}
have used these features to examine the star formation histories of
large samples of local galaxies from the Sloan Digital Sky Survey
(SDSS).  In the present paper we complement the
\citet{Kauffmannetal2003} work by measuring the same spectral features
in distant ($z=0.6-1.2$) galaxies from the Gemini Deep Deep Survey
\citep[GDDS, ][hereafter Paper I]{GDDS_PI}. Building two unbiased
samples of massive (\Msel{}) galaxies at $z=0.6-1.2$ and $z=0.05-0.1$
from the GDDS and the SDSS, respectively, we investigate the
differences in the star-formation histories at these two epochs.  Such
a comparison can shed light on the evolution of massive galaxies since
$z=2$: the local massive galaxies, dominated by elliptical
morphologies, are known to be nearly-passively evolving. Similarly
massive galaxies have also been found in surprisingly large numbers at
higher redshifts \citep[up to $z=2$, Paper III,][]{Fontanaetal2004} and some of these are already ``red and
dead'' \citep[hereafter Paper~IV]{Cimattietal2004,GDDS_PIV}. Our aim
is to use spectral features to better understand the detailed star
formation histories of these systems.

This paper is organized as follows. Our high-redshift and low-redshift
galaxy samples are described in Section~\ref{Section:surveys}.
Section~\ref{Section:indicators} presents the measurements we have
obtained from these data. Results obtained from these measurements are
given in Section~\ref{Section:analysis}. Modeling and interpretation
of our data are presented in Section~\ref{Section:models}, and our
conclusions are given in Section~\ref{Section:conclusion}.
Cosmological parameters $H_0=75$~km\,s$^{-1}$\,Mpc$^{-1}$,
$\Omega_\Lambda=0.7$, and $\Omega_M=0.3$ are assumed throughout this
paper.

%%%%%%%%%%%%%%%%%%%%%%%%%%%%%%%%%%%%%%%%%%%%%%%%%%%%%%%%%%%%%%%%%%%%%%%%%%%%%%%%%
\section{Sample Selection}
\label{Section:surveys}

The present study is based on two spectroscopic surveys at low and
high redshifts. At $<z>=1.1$, the GDDS contains more than 300
high-quality galaxy spectra and has a very well-understood selection
function (Paper I) designed to make it possible to use it as an unbiased
mass-limited sample. At low redshift ($z\simeq 0.1$), the SDSS provides many thousands
of galaxy spectra with excellent quality. In this Section, we first
give a brief overview of the common selection criteria for these two samples,
before describing the details of each sample in more detail.

\subsection{Overview}
\label{section:overview}
Our overall sample is selected from the GDDS and SDSS using three
criteria.  (1) We restrict our analysis to galaxies with secure
redshifts and high quality spectra. Our threshold corresponds to a
typical signal-to-noise ratio of $S/N > 2$ per pixel. (2) Then,
we restrict the redshift range of each sample to the maximum interval
over which the measurement of both \Hdelta{} and the 4000-\AA{} break
is possible in the data.  (3) Finally, we select galaxies with a total
stellar mass larger than $10^{10.2}$~M$_\odot$.  With this purely
physical selection we avoid color selection biases (for instance,
UV-selected samples can over-represent star-forming galaxies with
respect to red galaxies).  We choose this particular  mass cut-off
because it corresponds to the lowest mass for which the GDDS sample
can be considered as complete in the redshift range defined by our
second criterion. Doing so, we also maximize the number of galaxies
included in the high redshift sample, which makes our results more
robust. Finally, given this mass cut-off, we further restrict the
SDSS sample to the redshift range over which the survey is complete.
Eventually, we obtain two comparable, mass-limited samples at low
($0.05<z<0.1$) and high ($0.6<z<1.2$) redshifts.

\subsection{The GDDS sample}
%\onecolumn
\begin{table*}
\caption{Properties of our sample of GDDS massive galaxies.\label{table:props}}
\tablewidth{0pt}
\tabletypesize{\scriptsize}
\centering
\begin{tabular}{@{\extracolsep{5pt}}l l  r@{\extracolsep{0pt}\hspace{2pt}}l@{\extracolsep{5pt}} r@{\extracolsep{0pt}\hspace{2pt}}l@{\extracolsep{5pt}}  r@{\extracolsep{0pt}\hspace{2pt}}l@{\extracolsep{5pt}}  r@{\extracolsep{0pt}\hspace{2pt}}l@{\extracolsep{5pt}}  c c c c}
\hline
\hline
  \multicolumn{1}{c}{ID}   &  \multicolumn{1}{c}{z}   &        \multicolumn{2}{c}{\HdeltaA}   &    \multicolumn{2}{c}{\OIIEW} &  \multicolumn{2}{c}{\Dn{}} & \multicolumn{2}{c}{$\log_{10}(M_\ast/M_\odot)$}  & $g-r$ & SFR([OII]) & M$_\ast$/SFR & weight \\
     &          &      \multicolumn{2}{c}{(\AA{})}   &    \multicolumn{2}{c}{(\AA{})} &          &          &     &    & & ($M_\odot$yr$^{-1}$) & (Gyr) & \\
\hline
        GDDS-22-2863 &  0.918 & $ 11.81$ & $\pm   4.10$ & $-47.58$ & $\pm   8.72$ & $  1.16$ & $\pm   0.11$ & $ 10.30$ & $\pm   0.30$ & $  0.62$ & $   1.7$ & $    11.6$ &  0.070\\
        GDDS-22-0128 &  1.024 & $ 11.62$ & $\pm   2.65$ & $-56.41$ & $\pm  12.69$ & $  1.49$ & $\pm   0.13$ & $ 10.55$ & $\pm   0.28$ & $  0.60$ & $   1.7$ & $    20.7$ &  0.028\\
        GDDS-12-8139 &  1.189 & $  8.36$ & $\pm   0.54$ & $ -6.21$ & $\pm   0.65$ & $  1.61$ & $\pm   0.02$ & $ 10.39$ & $\pm   0.29$ & $  0.40$ & $   2.3$ & $    10.5$ &  0.019\\
        GDDS-02-1724 &  0.996 & $  7.58$ & $\pm   0.19$ & $ -2.45$ & $\pm   0.32$ & $  1.34$ & $\pm   0.01$ & $ 10.87$ & $\pm   0.09$ & $  0.37$ & $   2.6$ & $    27.7$ &  0.007\\
        GDDS-12-5722 &  0.841 & $  6.14$ & $\pm   0.58$ & $ -8.74$ & $\pm   1.11$ & $  1.34$ & $\pm   0.02$ & $ 10.69$ & $\pm   0.10$ & $  0.67$ & $   1.6$ & $    31.1$ &  0.166\\
        GDDS-22-0281 &  1.022 & $  6.10$ & $\pm   1.26$ & $-14.88$ & $\pm   2.31$ & $  1.60$ & $\pm   0.06$ & $ 11.02$ & $\pm   0.13$ & $  0.71$ & $   1.5$ & $    70.6$ &  0.166\\
        GDDS-02-1777 &  0.982 & $  4.74$ & $\pm   0.90$ & $-10.81$ & $\pm   1.71$ & $  1.31$ & $\pm   0.03$ & $ 10.43$ & $\pm   0.22$ & $  0.56$ & $   2.3$ & $    11.5$ &  0.033\\
        GDDS-02-1702 &  1.052 & $  4.58$ & $\pm   1.58$ & $-42.99$ & $\pm   2.42$ & $  1.17$ & $\pm   0.04$ & $ 10.40$ & $\pm   0.18$ & $  0.33$ & $   9.0$ & $     2.8$ &  0.020\\
        GDDS-02-1543 &  1.131 & $  4.22$ & $\pm   2.65$ & $ -5.34$ & $\pm   2.71$ & $  1.28$ & $\pm   0.06$ & $ 10.80$ & $\pm   0.15$ & $  0.67$ & $   0.2$ & $   267.4$ &  0.070\\
        GDDS-22-2548 &  1.022 & $  3.47$ & $\pm   1.51$ & $-17.08$ & $\pm   2.46$ & $  1.71$ & $\pm   0.07$ & $ 10.99$ & $\pm   0.18$ & $  0.70$ & $   2.3$ & $    41.5$ &  0.166\\
        GDDS-22-0315 &  0.909 & $  3.13$ & $\pm   0.96$ & $-14.90$ & $\pm   1.05$ & $  1.56$ & $\pm   0.03$ & $ 10.69$ & $\pm   0.10$ & $  0.53$ & $   4.2$ & $    11.7$ &  0.013\\
        GDDS-12-7660 &  0.791 & $  3.04$ & $\pm   0.79$ & $-15.66$ & $\pm   1.11$ & $  1.20$ & $\pm   0.02$ & $ 10.48$ & $\pm   0.14$ & $  0.46$ & $   2.5$ & $    12.3$ &  0.013\\
        GDDS-02-0715 &  1.133 & $  3.01$ & $\pm   2.14$ & $ -4.29$ & $\pm   2.22$ & $  1.35$ & $\pm   0.06$ & $ 11.25$ & $\pm   0.06$ & $  0.37$ & $   1.8$ & $   101.6$ &  0.053\\
        GDDS-15-6851 &  1.126 & $  3.00$ & $\pm   0.92$ & $ -8.32$ & $\pm   1.51$ & $  1.64$ & $\pm   0.03$ & $ 10.68$ & $\pm   0.13$ & $  0.62$ & $   4.2$ & $    11.5$ &  0.028\\
        GDDS-12-5337 &  0.679 & $  2.30$ & $\pm   0.39$ & $-13.30$ & $\pm   1.17$ & $  1.67$ & $\pm   0.02$ & $ 10.42$ & $\pm   0.13$ & $  0.50$ & $   1.9$ & $    14.0$ &  0.027\\
        GDDS-12-8983 &  0.963 & $  0.72$ & $\pm   0.34$ & $ -1.51$ & $\pm   0.50$ & $  1.77$ & $\pm   0.02$ & $ 10.70$ & $\pm   0.07$ & $  0.56$ & $   0.6$ & $    89.7$ &  0.015\\
        GDDS-22-0893 &  0.875 & $  0.34$ & $\pm   1.38$ & $ -6.52$ & $\pm   1.81$ & $  1.93$ & $\pm   0.07$ & $ 10.59$ & $\pm   0.18$ & $  0.68$ & $   0.6$ & $    60.2$ &  0.033\\
        GDDS-02-0857 &  1.049 & $  0.13$ & $\pm   1.23$ & $ -6.51$ & $\pm   1.28$ & $  1.88$ & $\pm   0.05$ & $ 11.21$ & $\pm   0.06$ & $  0.79$ & $   1.8$ & $    87.1$ &  0.168\\
        GDDS-12-6800 &  0.615 & $ -0.13$ & $\pm   0.34$ & $  0.95$ & $\pm   0.64$ & $  2.03$ & $\pm   0.02$ & $ 10.89$ & $\pm   0.09$ & $  0.73$ & $   0.1$ & $  1443.0$ &  0.038\\
        GDDS-12-6456 &  0.612 & $ -0.43$ & $\pm   0.57$ & $ -0.06$ & $\pm   1.08$ & $  1.99$ & $\pm   0.04$ & $ 10.72$ & $\pm   0.14$ & $  0.73$ & $   0.2$ & $   286.3$ &  0.027\\
        GDDS-02-1011 &  1.133 & $ -1.04$ & $\pm   2.00$ & $  4.93$ & $\pm   1.53$ & $  1.78$ & $\pm   0.07$ & $ 10.48$ & $\pm   0.13$ & $  0.65$ & $   0.3$ & $    88.8$ &  0.070\\
        GDDS-02-1935 &  0.915 & $ -1.12$ & $\pm   2.84$ & $ -4.28$ & $\pm   2.84$ & $  1.82$ & $\pm   0.10$ & $ 11.12$ & $\pm   0.07$ & $  0.77$ & $   0.4$ & $   336.7$ &  0.166\\
        GDDS-15-5348 &  0.964 & $ -1.97$ & $\pm   1.19$ & $-10.82$ & $\pm   3.55$ & $  1.96$ & $\pm   0.07$ & $ 11.06$ & $\pm   0.07$ & $  0.71$ & $   0.8$ & $   149.1$ &  0.166\\
        GDDS-15-7241 &  0.749 & $ -2.29$ & $\pm   0.68$ & $ -1.30$ & $\pm   0.98$ & $  1.74$ & $\pm   0.03$ & $ 10.70$ & $\pm   0.14$ & $  0.62$ & $   0.2$ & $   260.4$ &  0.011\\
        GDDS-22-0510 &  0.820 & $ -2.30$ & $\pm   1.13$ & $ -2.63$ & $\pm   1.34$ & $  1.81$ & $\pm   0.04$ & $ 10.51$ & $\pm   0.13$ & $  0.67$ & $   0.6$ & $    55.0$ &  0.013\\
\hline
\end{tabular}
\end{table*}

%\twocolumn
The characteristics of the Gemini Deep Deep Survey are described in
Paper I; only a very brief description of the survey is
given here. The GDDS targets galaxies in the ``redshift desert''
($1<z<2$). The survey contains 312 spectra of galaxies in four fields
of the Las Campanas Infrared Survey \citep{LCIRS2001}.  The primary
selection is based on near-infrared photometry ($K_s<20.6$, $I<24.5$
in Vega magnitudes) giving preference to galaxies with red $I-K$
colors and photometric redshifts above $z=1$
\citep{Chenetal2002,Firthetal2002}.

%spectroscopy
Spectra were obtained with the GMOS multi-slit spectrograph on the
Gemini North 8~meter telescope. The ``Nod \& Shuffle'' technique was
used to optimize sky subtraction in order to allow very long exposures
($\gtrsim 30$ hours per field) to be undertaken. This yielded
high-quality spectra with {\sc FWHM}$\simeq 17$~\AA{} from which
spectroscopic redshift determinations were successful for $\simeq
75$\% of the sample. About 80\% of the galaxies with secure redshift
also satisfy the condition $S/N>2$ per pixel around the \Hdelta{}
line.

In this paper our sample is restricted to the 62 galaxies with
high-quality spectra and secure redshifts in the range
$0.6<z<1.2$. This range corresponds to the maximum interval over which
all galaxies have measurable \HdeltaA{} and \Dn{} indices.  We further
limit the sample to galaxies with stellar masses \Msel{}, leading to a
subsample of 26 galaxies.  The choice of this mass cut-off maximizes the number of galaxies in our GDDS
sample (it is the minimum mass for which our sample is mass-complete
over the redshift range $0.6<z<1.2$, as demonstrated in
Paper~III). Stellar masses, derived from template fits to the
multicolor (VIK) photometry, are taken from Paper~III.  The masses used in
this paper assume a \citet{Baldry&Glazebrook2003} Initial Mass
Function (hereafter BG03 IMF). The typical error bar on
$\log_\mathrm{10} M_\star$, estimated by Monte Carlo simulations, is
of the order of 0.2~dex.  A single galaxy (GDDS-02-2197) is then
removed from this subsample because of very doubtful measured
\HdeltaA{} and \Dn{} indices.  This galaxy is an outlier with the
largest error bars in the sample.  In fact, these error bars are so
large that the object is compatible with the rest of the sample in a
formal statistical sense, so simply omitting it changes none of our
conclusions while simplifying the analysis. This leaves us with 25
galaxies in the final GDDS subsample.  The main properties of these 25
galaxies are given in Table~\ref{table:props}. A full description of
all the measured quantities present in this table is given in
Section~\ref{Section:indicators}.

\begin{figure*}[!tbf]
\centering
\includegraphics[height=18cm]{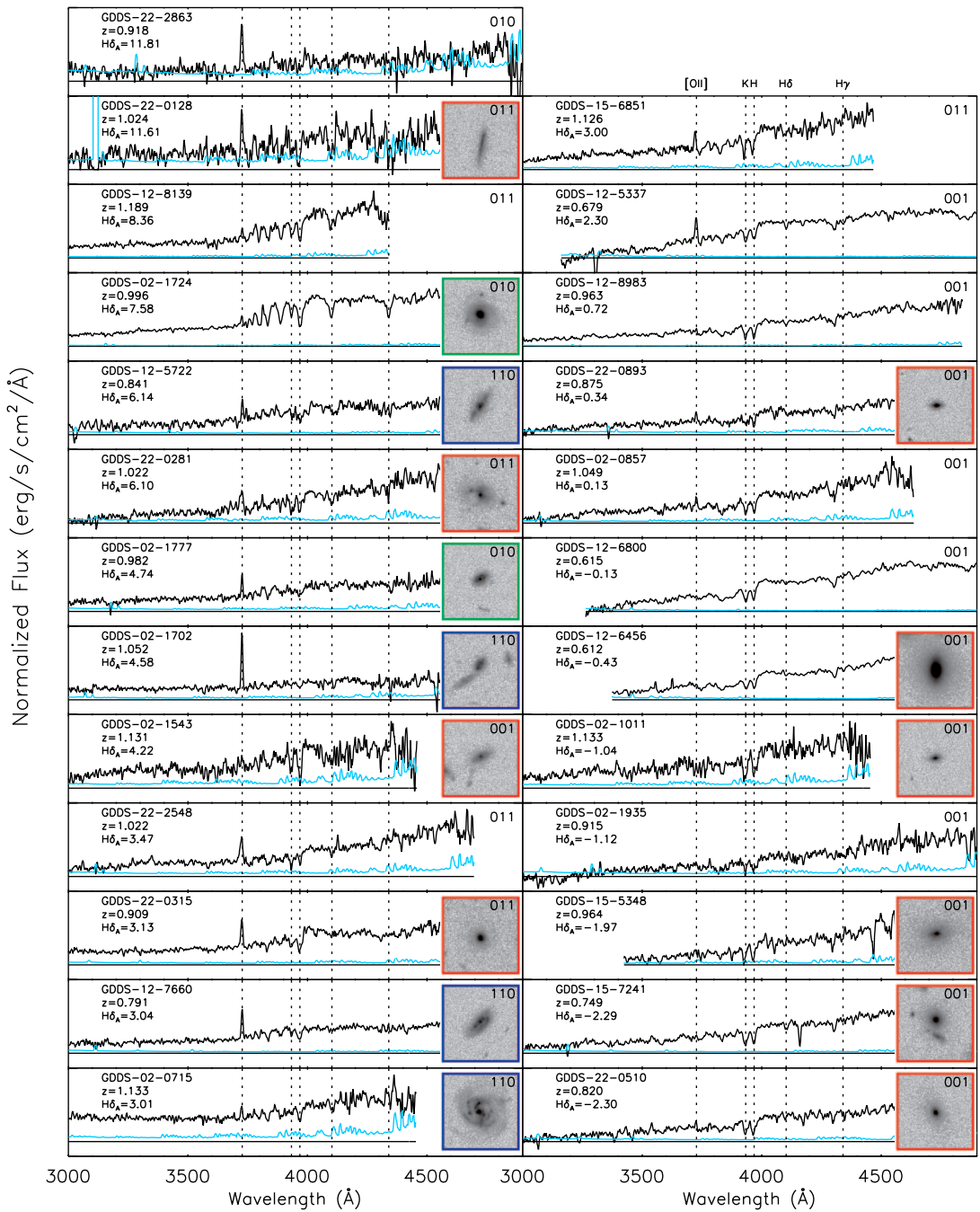}
\figcaption{Sample of the 25 mass selected GDDS galaxies, sorted by
decreasing \HdeltaA{} (expressed in \AA{}). The error spectra are in
blue and HST ACS images are shown when available. The size of these
images is 5x5 arcsec$^2$. The color of the frame around each image
codes the luminosity-mean age of the galaxy (estimated by visual
inspection of the spectrum and broad-band SED, cf Paper~I): red for old stellar
populations (codes 001 and 011), green for intermediate age (code 101)
and blue for young stellar populations (code 110). Note that slightly
negative values for \HdeltaA{} ($\lesssim 3$\AA{}) do not necessarily
correspond to emission lines (e.g. GDDS-15-7241), because of the very
definition of this Lick index: even when the \Hdelta{} line is in
absorption, the pseudo-continuum can be ``lower'' than the line
itself.
\label{figure:allspectra}}
\end{figure*}

Figure~\ref{figure:allspectra} presents rest-frame spectra of all 25
GDDS galaxies investigated in the present paper, with a 3-pixel
boxcar smoothing, together with the error spectra (in blue) and their
ACS HST images \citep[see also][]{GDDS_morpho} when available. The
color of the frame around each image codes the spectral classification of the galaxy
(Table 5 of Paper I, red for old stellar populations, green for
intermediate age and blue for young stellar populations).  The median
redshift of this sample is $\bar{z}=0.964$.

The selection function for the GDDS is given in Paper~I.  The inverse
weights in that paper quantify the degree to which each galaxy is
over-represented in a $I-K$ vs. $I$ color-magnitude diagram. It is important to note that the
primary goal of the GDDS survey was to study
red galaxies at high redshift. Therefore, by design, this survey
specifically targeted red galaxies with a sparse sampling rate of
$\sim$1/2 compared to $\sim$1/7 for the remainder of the photometric
sample. The inverse weights must therefore be used to account for this unbalance. As we have
just described in Section~\ref{section:overview}, in the present paper further restrictions on the
sample are introduced (so as to only include galaxies with high
confidence redshift class and high spectral quality), and a new set of
weights is needed to define the selection function for this
sample. These are given in the last column
Table~\ref{table:props}\footnote{We emphasize that these are {\em
inverse} weights. A small number implies that the object needs to be
weighted more heavily because it is under-represented in the sample.}.
In Section~\ref{Section:analysis}, we will show that our results are
not strongly dependent on the adopted weights.

%\clearpage

%\clearpage

\subsection{The SDSS sample ($\bar{z}\simeq 0.1$)}
At low redshift, we define a sample of galaxies drawn from the Sloan
Digital Sky Survey DR2 \citep{SDSSDR2}. We restrict
the present investigation to the `main' galaxy sample, i.e. galaxies
with Petrosian $r$-band magnitudes in the range $14.5<r<17.77$.
Only galaxies with secure redshift measurements are considered. To
reduce sampling incompleteness, we focus on galaxies with $z>0.05$,
and in order to be consistent with the quality of the GDDS spectra, we
also exclude the few galaxies (0.2\% of the sample) with $S/N<2$.

As for the GDDS sample, we also impose a mass cut, so that our sample
includes only massive galaxies (\Msel{}). According to the SDSS
stellar mass function determined by \citet{Panteretal2004}, the sample
is complete for $\log_{10}(M_\star/h^{-2}) \geq 10.2$ only at
$z<0.1$. Using the cosmology adopted in our study, this corresponds to
$M_\star \geq 10^{9.9} M_\odot$. However, this mass-function is
estimated using a \citet{Salpeter1955} IMF, while the stellar masses
of GDDS galaxies are estimated using the BG03 IMF. The conversion
factor between these mass estimates is $M_\mathrm{SP} = 1.82
M_\mathrm{BG}$ (Paper~III), or 0.26 dex. Including this correction, we
consider that the SDSS sample is complete for \Msel{} (using the BG03
IMF) for $z<0.1$.  We use the mass estimates from
\citet{Kauffmannetal2003} which are corrected from
dust-obscuration. Because these authors use a \citet{Kroupa2001} IMF,
we subtract 0.07~dex from their masses to convert to BG03 IMF mass
estimates. For our subsample of SDSS galaxies, the typical 95\%
confidence range on $\log M_\star$ is 0.2 dex.

After imposing the cuts described above, our final SDSS low-redshift sample
contains $50,255$ galaxies.

%%%%%%%%%%%%%%%%%%%%%%%%%%%%%%%%%%%%%%%%%%%%%%%%%%%%%%%%%%%%%%%%%%%%%%%%%%%%%%%%%
\section{Measurements}
\label{Section:indicators}

In this Section, we provide a detailed description of
the quantities measured from our low-redshift
and high-redshift samples. The reader who is
not interested in details of how
these measurements were made may wish to
simply note that we measured a single rest-frame
broad-band color and the spectral indices
listed in Table~\ref{table:indices}, and then move ahead to 
Section~\ref{Section:models}.

\subsection{Rest-frame g-r color}
At low redshift, SDSS galaxy spectra and fiber
u,g,r,i,z magnitudes are publicly available.
However, given that we use the SDSS spectra to measure other spectral
indices (\Dn{} and \HdeltaA{}), we decided to measure a color directly
on the SDSS spectra to be consistent in our measurements
(and to avoid, for example, aperture effects which might affect differently the spectral features and the photometry).
Because of this choice, $g-r$ and $B-V$ are the two main optical colors
available for the SDSS galaxies. We checked that both colors give
similar results, and we concentrate on the $g-r$ color in the
following. The $g-r$ color is measurable on 95\% of the SDSS spectra.

Unfortunately, because of the limited wavelength coverage of the GDDS
galaxy spectra, it is not possible to measure a standard optical color
from them in the same way.  At redshift~$\simeq 1.1$, the optical
spectra of the GDDS galaxies correspond to UV rest-frame
spectra. However, the existence of optical and near-infrared
photometry makes it possible to estimate a rest-frame optical color,
using k-corrections, in the following way.  The rest-frame $g-r$ color
of the GDDS galaxies was measured on the synthetic spectra that fit
best the observed spectral energy distributions, using the
B,V,R,I,z,J,H and K photometry.  The procedure used to fit the data
combines two instantaneous bursts of star formation at different ages
and metallicities, reddened by dust (parameterized by \ebv{}, ranging
from 0 to 2). A minimization of the $\chi^2$ using the ages, the
metallicities and the extinction as free parameters, provides a
best-fitting synthetic spectrum.  (It is worth mentioning that this
procedure was not used to derive ages, metallicities, or extinction
parameters: doing so would require a much more refined grid than the
one we used because of the degeneracies that exist in the measurement
of these three parameters.)  Column~8 of Table~\ref{table:props} lists
the measurements of the $g-r$ color for the GDDS galaxies in our
sample.

\subsection{\Dn{}}
\begin{table}
\caption{Definitions of the spectral indices used in this paper.
\label{table:indices}}
\tablewidth{0pt}
\startdata
\centering
\begin{tabular}{lccc}
\hline
\hline
            &      Blue band     & Feature band    & Red band \\
Index        &      (\AA)       &  (\AA)    &  (\AA) \\
\hline
\Dn{}       &      3850.00-3950.00&                &    4000.00-4100.00 \\
\HdeltaA{}  &      4042.85-4081.00 & 4084.75-4123.50   & 4129.75-4162.25\\
\HdeltaEW{} &      4017.00-4057.00 & 4083.50-4122.25 &  4153.00-4193.00 \\
\OIIEW{}    &      3696.30-3716.30  &  3716.30-3738.30   & 3738.30-3758.30 \\
\hline
\end{tabular}
\end{table}

The $g-r$ color of a galaxy can be affected by the presence of dust,
and the \Dn{} at 4000~\AA{} is often used to estimate ages instead because
it is much less sensitive to extinction. For example, a reddening of
\ebv{}=0.5 shifts \Dn{} by less than 10 percent, which is of the
order of the change produced by a relative difference in age of 0.25
dex (or $\simeq 45$\%).  This effect is small in comparison to the
corresponding reddening of the $g-r$ color, which can be mistaken for
a $\simeq 1$~dex age effect.  Remarkably, the dependence of \Dn{} on
the metallicity is much stronger than its dependence on the
extinction: for a 10~Gyr old population, it varies from 1.7 to 2.7
when the metallicity varies from $Z_\odot / 5$ to $2.5 Z_\odot$
\citep[e.g.][]{Kauffmannetal2003}.  The definition of \Dn{}
\citep[from][]{Baloghetal1999} is reported in
Table~\ref{table:indices} together with the other indices used in this
paper.

\subsection{The \Hdelta{} line}

In the absence of emission, the equivalent widths of Balmer absorption
lines decrease steadily with time for an old (age$>1$~Gyr) passively
evolving stellar population. Moreover, they are quite generally little to
the mean stellar metallicity and dust content of a galaxy.  However,
in practice, H$\alpha$ and H$\beta$ are often partially filled in by
nebular emission, making their interpretation more difficult. Very
high order lines are generally too weak to be easily measurable, so
H$\gamma$ and \Hdelta{} are the two lines of choice for probing
stellar ages. In the present study we concentrate on \Hdelta{}
because, unlike the lower order Balmer lines, it is measurable in
optical spectra of galaxies redshifted up to $z=1.2$.

An interesting property of the \Hdelta{} absorption line lies in its
behavior between the birth of a generation of stars and the age of
2~Gyr: the equivalent width first increases strongly during a few
$10^8$~years, reaches its maximum value (almost 10~\AA{}) when the population is dominated
by A stars, and then decreases continuously, passing through its
initial value when the population is about 2~Gyr old. This behavior
implies that high equivalent widths in absorption ($\gtrsim 4$~\AA{})
can only be reached in galaxies where some substantial star formation
occurred a few $10^8$~years before, and where the subsequent star
formation events were weak.  We will refer to systems with ${\rm
H}\delta_A>4$~\AA~ as being ``H$\delta$-strong'' (HDS).

To measure the equivalent width of \Hdelta{}, we use the definition of
the Lick index \HdeltaA{} \citep{Worthey&Ottaviani1997} which is well
suited to probe stellar populations dominated by A stars in the
optical.  We degrade the resolution of all the rest-frame spectra to
the appropriate Lick resolution (11~\AA), following the prescription
of \citet{Worthey&Ottaviani1997}. 

\begin{figure}[!tbf]
\plotone{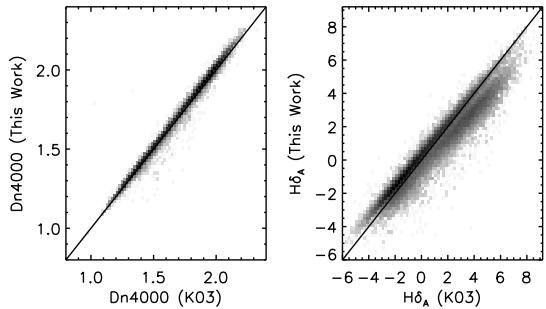} 
\figcaption{Comparison of \Dn{} and \HdeltaA{} measurements on our SDSS
sample between \citet{Kauffmannetal2003} (with a correction for
nebular emission and a higher spectral resolution) and this work
(including the emission lines and at the Lick spectral
resolution). The grey scale level represent the number of galaxies in
each pixel in a logarithmic scale,  darker pixels having more galaxies.
\label{figure:comparison_K03}}
\end{figure}

For the sake of consistency between the two data sets, we choose to
measure the \HdeltaA{} and \Dn{} indices ourselves on the SDSS
spectra, rather than using the previously published values.  To
validate these measurements, we compare them with the values published
by \citet[][hereafter K03]{Kauffmannetal2003} in
Fig.~\ref{figure:comparison_K03}.  Perfect agreement is not expected:
the differences between our measurements of \HdeltaA{} and the values
given by K03 can be partly explained by the different spectral
resolutions ($\simeq 3$~\AA{} for K03, and $\simeq 11$~\AA{} for this
work which uses the Lick resolution).  Moreover, unlike those authors,
we choose not to correct \HdeltaA{} from the nebular emission for two
reasons: first, we prefer to compare the raw data to models that
include the contribution from emission lines; second, the correction
from the contribution from emission lines to the \HdeltaA{} index is
too uncertain for the GDDS galaxies which have on average lower $S/N$
ratios than the SDSS spectra. However, the agreement between the two
measurements is good, given the different considerations involved in
the measurements. For large values of \HdeltaA{}, our measured values
are smaller than the K03 values by $\simeq
1$~\AA{}.  This is to be expected as a correction for line emission in
such young populations would increase \HdeltaA{} by about $1\--3$
\AA{}, as predicted by the {\sc P\'egase} models. A similar correction
is mentioned in \citet{Baloghetal1999}, and is confirmed by
\citet{Gotoetal2003} who measured a median correction of 15\% for the
\Hdelta{} equivalent width on a sample of galaxies drawn from the
SDSS.

%\clearpage
%\clearpage

\subsection{Emission lines}
The nebular emission lines produced in \ion{H}{2} regions are excellent
indicators of recent star-formation. In the redshift range of the GDDS
sample, the main visible lines are [OII]$\lambda3727$ and the \Hdelta{}
Balmer line (H$\gamma$ and H$\beta$ are too much redshifted to be visible). In
the following, we concentrate on the equivalent width of the
[OII]$\lambda3727$ line as an indicator of ongoing or very recent
(less than 50~Myr) star formation.

\subsection{Errors}
\label{section:errors}
The uncertainties on the measurements of \HdeltaA{} and \Dn{}, and
\OIIEW{}, reported in Table~\ref{table:props}, are computed following
\citet{Cardieletal1998}, using, for each galaxy, the measured mean
signal-to-noise ratios inside the integration bands of
Table~\ref{table:indices}. To confirm these error bars,
Monte-Carlo simulations were carried out assuming a Gaussian
distribution for the error spectrum. For each galaxy, 50 simulations
were considered, and the rms dispersion of the
distribution of indices matches very well the error bars
predicted using the formulae from
\citet{Cardieletal1998}: the rms for the distribution of relative
differences between these two errors is 0.22 for \HdeltaA{} and 0.12
for \Dn{}.

Although we believe that the observed  fractions of HDS galaxies in each redshift
range are robust, several potential sources of
errors should first be considered.

The very definitions of the indices used might be a source of
systematic errors. We confirmed that the use of the \HdeltaEW{}
definition instead of  the index \HdeltaA{} does not
change any of our conclusions.

Some errors may also come from the calibration of the original
data. In particular, the reddest end of the GDDS spectra is known to
be subject to a  quite uncertain correction (the "redfix''
correction, cf. Paper I) due to charge diffusion in the CCD detectors
in the red. Fortunately, this correction applies to the portion of the
rest-frame spectrum which contains the \Hdelta{} line only at
redshifts larger than $1.3$. Therefore, it does not affect our study.
The selection function, which introduces important weighting, might
also be a concern. However, our results are qualitatively the same if
this selection function is omitted.  Finally, we estimate that the
aperture effects, which could potentially affect the observed
integrated spectral indices and colors, are small on average. Indeed, as GDDS
galaxies are observed at higher and higher redshift, a larger fraction
of the galaxies in enclosed in the 0.75'' aperture slit (corresponding
to a physical diameter of 4.7~kpc at $z=0.6$ and 5.8~kpc at $z=1.2$),
and more light from the outer parts contributes to the integrated
indices. However, over the whole $0.6<z<1.2$ interval, a larger
fraction of the total light is measured than that measured by the SDSS
fibers on the $z=0.05-0.1$ galaxies. The recent work of
\citet{Brinchmannetal2004} and \citet{Kewleyetal2004} showed that
aperture effects for emission-line-derived SFR in the SDSS galaxies is
negligible for $z>0.05$. The aperture effects on the stellar
absorption lines (typically more concentrated than the emission-lines)
are expected to be negligible.

Overall, the main sources of uncertainties are unlikely to explain the
observed trend of the decrease of the fraction of HDS galaxies since
$z=1.2$ described in next Section. We conclude that this evolution
must be real and can only be explained by changes in the star
formation histories of massive galaxies during the past 10~Gyr.

%%%%%%%%%%%%%%%%%%%%%%%%%%%%%%%%%%%%%%%%%%%%%%%%%%%%%%%%%%%%%%%%%%%%%%%%%%%%%%%%%
\section{Compared properties of local and distant massive galaxies}
\label{Section:analysis}
\subsection{General results}
\label{Section:zdists}

%\clearpage
\begin{figure*}[!tbf]
  \centering
\includegraphics[width=0.8\textwidth]{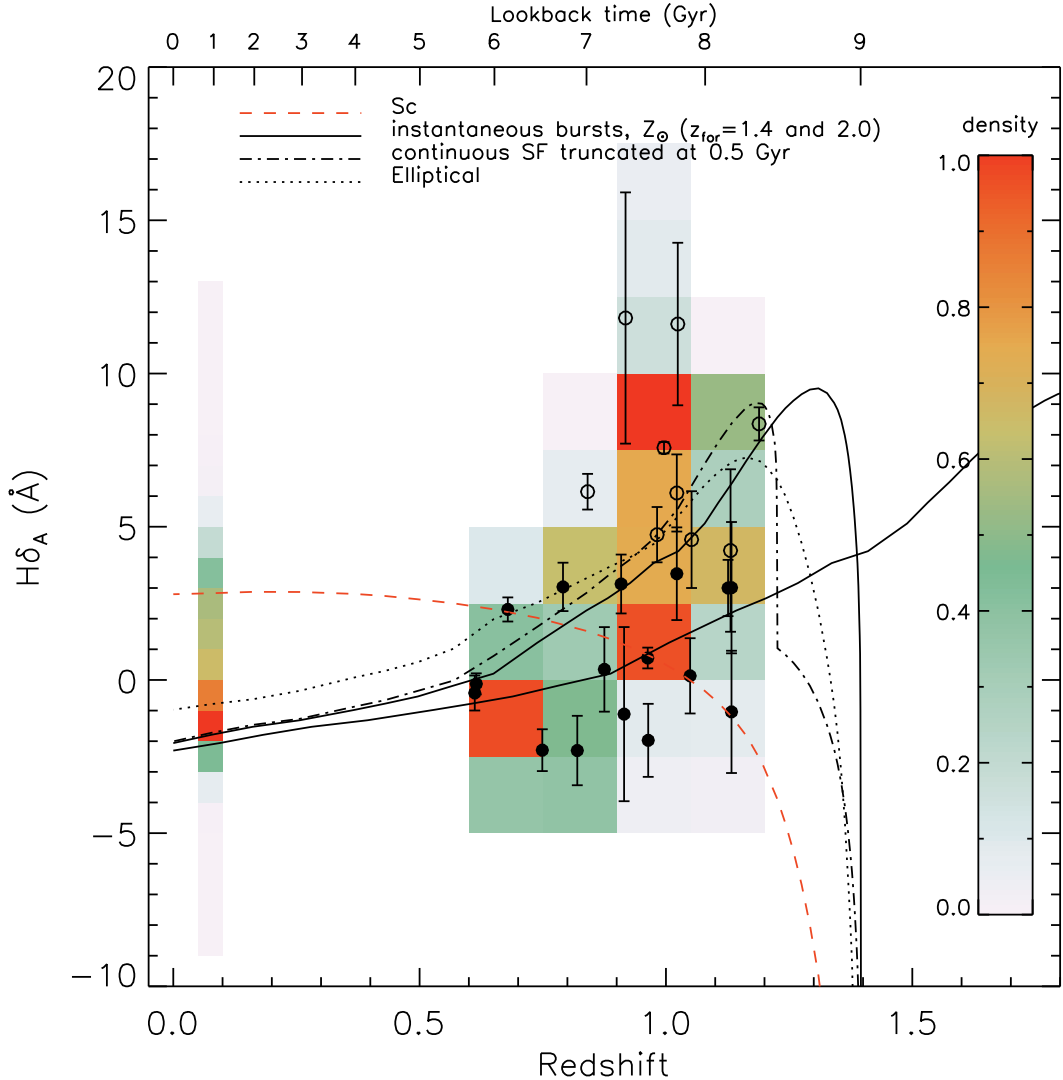}
\figcaption{Evolution of the \HdeltaA{} index from $z=1.2$ (GDDS sample)
  to the present (SDSS sample). For each sample, the plane is gridded
  and the number of galaxies that can be counted in each cell (we call
  this quantity ``density'') is color-coded: a density of 1 (reddest
  or darkest color) is assigned to the cell containing the most
  objects. The GDDS densities are converted from raw number counts by
  using the selection function and Monte Carlo simulations. The GDDS
  individual galaxies are represented by filled (\HdeltaA{}$\le
  4$~\AA{}) and empty (\HdeltaA{}$> 4$~\AA{}) circles. The models
  described in Section~\ref{Section:models} are over-plotted: ``burst''
  mode models are represented by black lines and ``quiescent'' mode
  models by red lines. The two 'cycle' models plotted in the following
  figures are omitted here for clarity: they would blur the figure,
  because their \HdeltaA{} indices alternatively reach very positive
  (when star-formation has ceased) and very negative values (because
  of emission lines) on short time-scales.
\label{figure:z_Hdelta}}
\end{figure*}

\begin{figure*}[!tbf]
\includegraphics[width=0.5\textwidth]{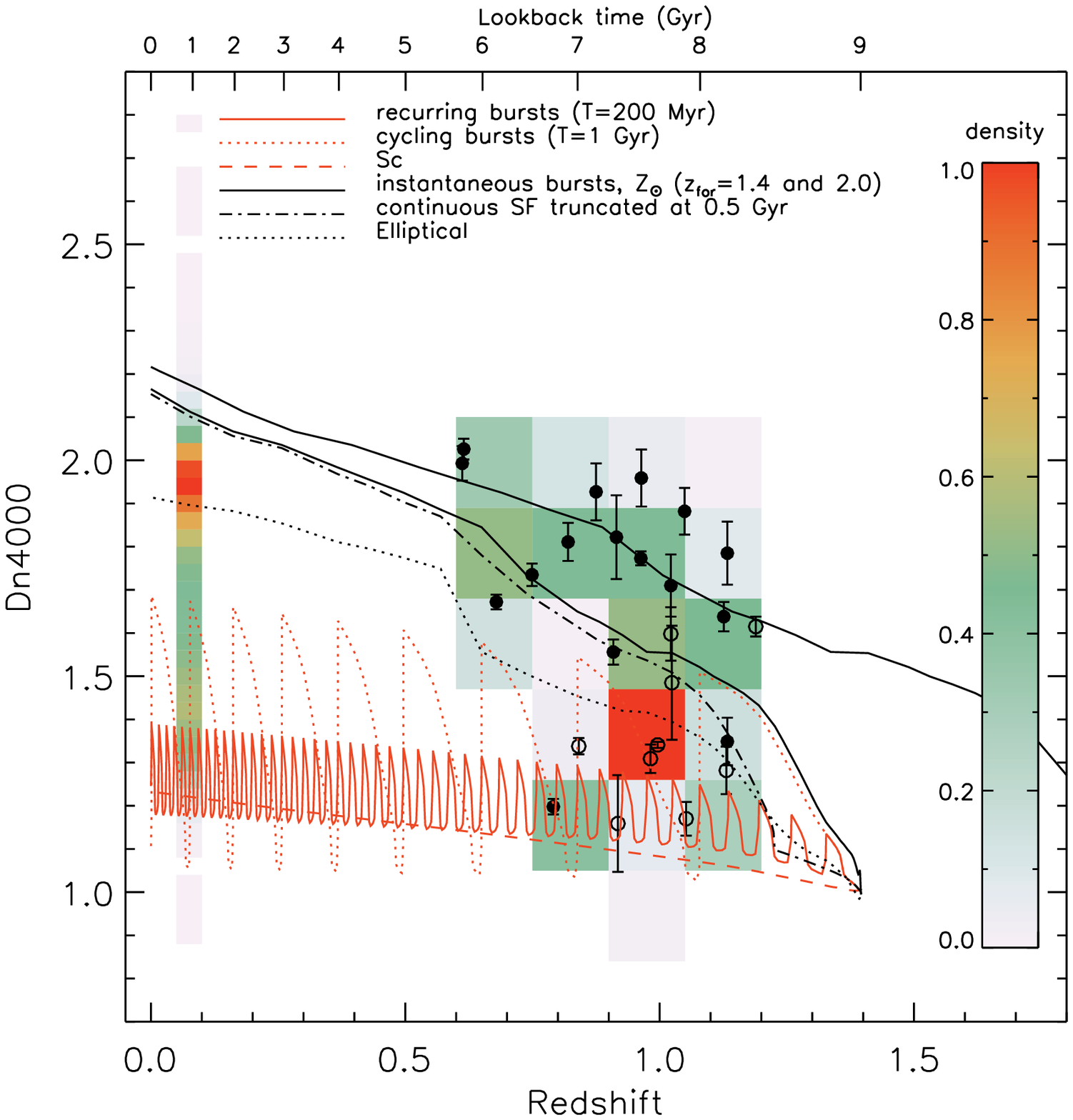}
% with cycles
\includegraphics[width=0.5\textwidth]{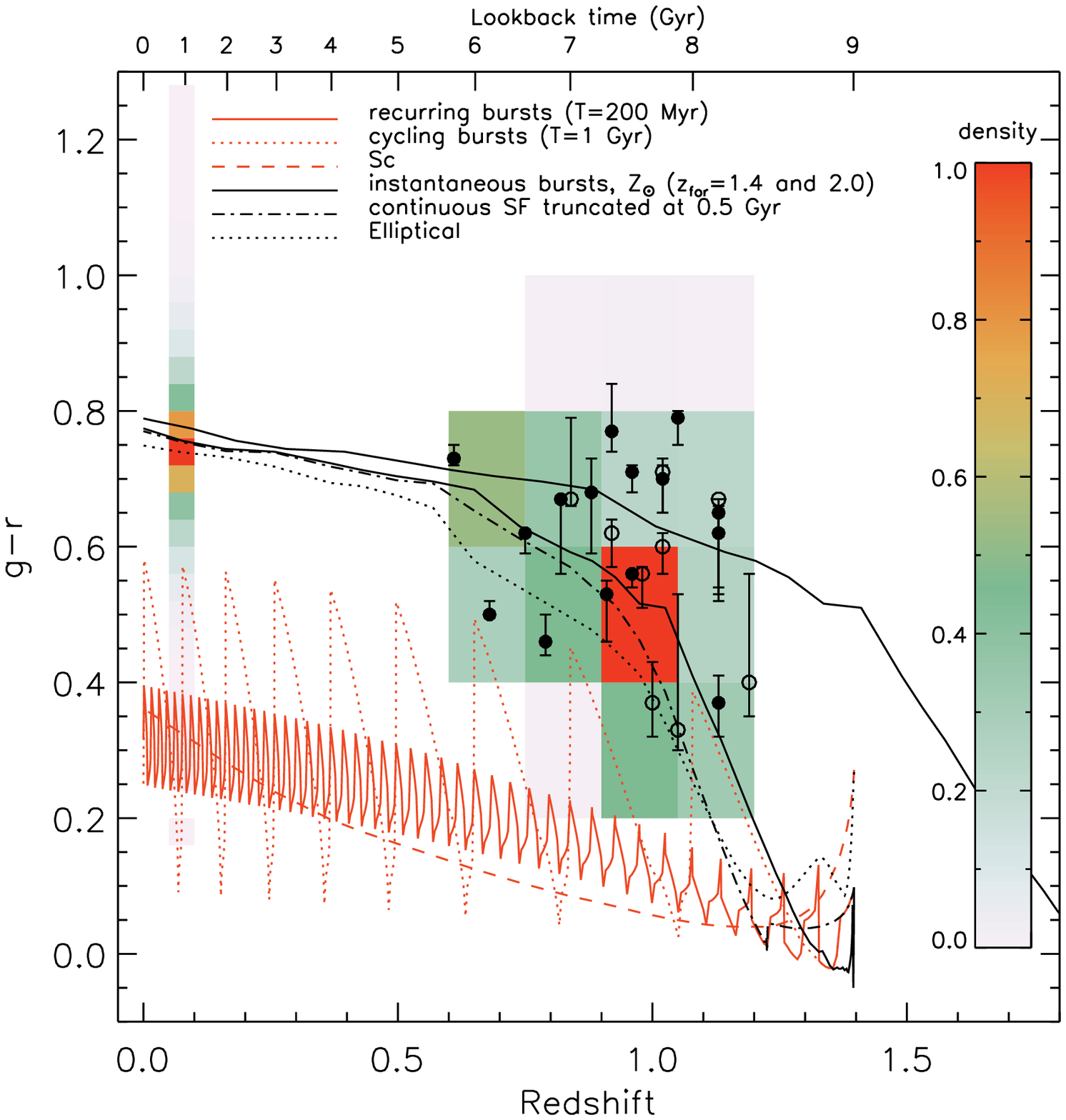}
\hfill% rest frame colors vs z with cycles
\figcaption{Similar to Fig.~\ref{figure:z_Hdelta} but for two other tracers of the
  luminosity-weighted age of galaxies: \Dn{} and the $g-r$ rest-frame
  color. Symbols as in Fig.~\ref{figure:z_Hdelta}.
\label{figure:zdists}}
\end{figure*}
%\clearpage

Measurements of the colors and spectral features in the galaxies
in our GDDS sample are presented in Table~\ref{table:props}.
Figures~\ref{figure:z_Hdelta} and \ref{figure:zdists} show the
distributions of \HdeltaA{}, \Dn{}, and $g-r$ rest-frame color as a function of
redshift. In these figures,  GDDS galaxies 
categorized as H$\delta$-strong (\HdeltaA{}$\ge 4$~\AA{}) are denoted
with open circles, and the remaining galaxies of the GDDS
sample are shown as filled circles. In addition, these figures show
two-dimensionally binned distributions (for both the GDDS sample and
the SDSS sample) using colored cells. The reddest (or darkest) cells
contain the largest number of galaxies. The size of the cells is
chosen to be on the order of, or larger than, the typical error bar of
the various observed quantities. The density of GDDS galaxies in each
binned cell has been corrected using the weights given in
Table~\ref{table:props}, and  Monte-Carlo simulations of each GDDS
galaxy were used to estimate the true distributions shown
in these figures.  (Note that this explains why the spread data in the
color-coded cells is a little larger than the coverage of the
individual points). 
Each of these figures also shows evolutionary models 
which will be described in the next section.

A comparison of the distributions of the GDDS and SDSS samples in
these figures shows a number of interesting results. The most striking
difference between the low-redshift and the high-redshift samples is
the disproportionately large number of H$\delta$-strong systems at
$z\sim 1$ in the GDDS sample, clearly visible in Figure~\ref{figure:z_Hdelta}.

\begin{figure}
  \centering \includegraphics[width=0.49\textwidth]{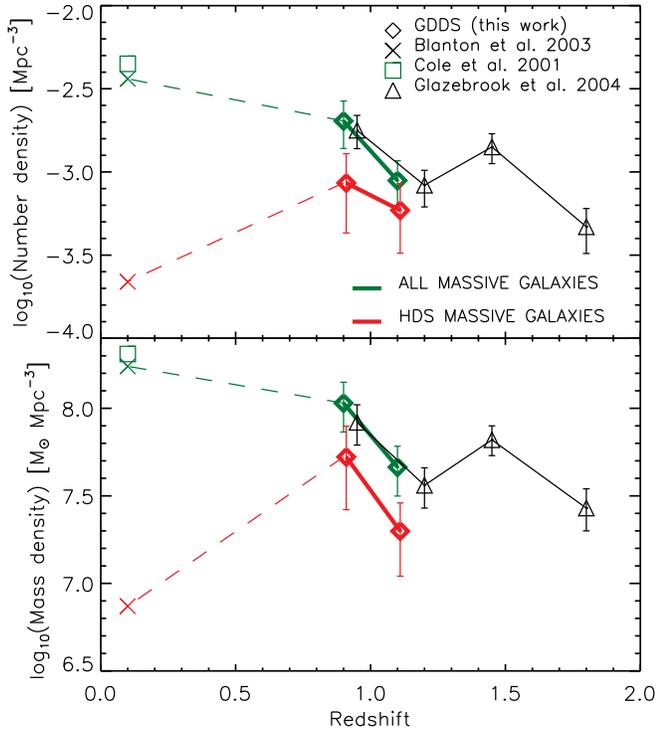}
  \figcaption{Evolution of the number and mass densities of massive
  galaxies and massive HDS (\HdeltaA{}$>4$\AA) galaxies. All these
  measurements concern only galaxies more massive than
  \Msel{}. Sampling weights are applied to GDDS galaxies, and the
  $1/V_{max}$ formalism is used at every redshift. Poisson error bars
  are shown for the $z \approx 1$ GDDS galaxies. These are too small
  to be shown for the $z \approx 0$ samples. Densities at low redshift
  were taken from \citet{Blantonetal2003} (from SDSS data, crosses)
  and \citet{Coleetal2001} (from 2dF data, squares), and corrected for
  the same IMF and cosmology. Mass densities from Paper~III
  \citep{GDDS_PIII} are shown in the lower panel for reference with a
  black solid line; the corresponding number densities are shown in
  the upper panel. The high redshift (GDDS) and low redshift (SDSS)
  samples used in this study are linked with dashed lines.
 \label{figure:densities}}
\end{figure}

To quantify this observation, we measure the number space densities
(per comoving Mpc$^3$) of massive galaxies and massive HDS galaxies in
various redshift intervals. For GDDS galaxies ($0.8\le z<1$ and $1 \le
z<1.2$), we use the sampling weights defined above and the $1/V_{max}$
formalism \citep{Schmidt1968} to correct from the fact that the
faintest galaxies in our sample could not be observed if they were at
higher redshift. At low redshift, we use the SDSS sample and the
$V_{max}$ values from \citet{Blantonetal2003}. At every redshift, we
treat separately the population of HDS galaxies, defined by
\HdeltaA{}$ \ge 4$~\AA{}. Figure~\ref{figure:densities} shows the
evolution of the number densities and mass densities from $z\approx 1$
to $z\approx 0$ for massive galaxies (\Msel{} with the BG03 IMF). In
particular, it shows that the $z\approx 1$ massive galaxies are about
one third (in number) of the local massive galaxies. This means that
the GDDS galaxies in our sample are almost certainly progenitors of
some of the local massive galaxies (because the stellar mass is
expected to either increase or keep steady), but about two thirds of
the local massive galaxies had progenitors at $z \approx 1$ that were
not massive enough to pass the mass threshold of \Msel{}. The figure
also confirms previous results \citep{GDDS_PIII,Fontanaetal2004},
showing that the total stellar mass contained in massive galaxies at in $z \approx 1$ is
already about a third of its present-day value. It
is more difficult to link this result with the claim of
\citet{Belletal2004} that the stellar mass in red galaxies must have
doubled since $z=1$. Indeed, our mass-selected sample contains not
only galaxies on the red-sequence, but also star-forming
galaxies (see Table~\ref{table:props}). Moreover, the mass cut-off adopted here probably excludes
the faint-end of the red-sequence at $z \approx 1$.

It also appears for the first time that although the number and mass
densities of all massive galaxies both increase by a factor of 2-3
from $z=1$ to $z=0$, these densities follow the opposite trend for HDS
massive systems: they actually decrease by a factor of 2-4 (for number
densities) and 2-7 (for mass densities) during the same period.  As a
result, the fraction of massive HDS galaxies is indeed much higher at
$z\approx 1$ ($\simeq1/2$) than at $z\approx 0$ (less than $1/10$). A
direct consequence of this strong decrease of number densities for HDS
galaxies is that most of the $z=1$ HDS massive galaxies necessarily
end up in local non-HDS (and probably passively evolving)
galaxies. This disconnect between distant and local HDS galaxies is
not too surprising because the timescales involved are quite
different: less than 1-2 Gyr for the HDS phase and $\sim 7$~Gyr for
the Hubble time interval between these two redshifts.

%\clearpage
%\clearpage

In the second part of this section we will consider how best to parameterize this
evolution, and explore the extent to which the evolution depends on
the equivalent width threshold beyond which galaxies become flagged as
HDS systems. 

The comparison of SDSS and GDDS samples in Figure~\ref{figure:zdists}
shows complementary information. The non-HDS GDDS galaxies (black
dots) tend to have \Dn{} indices almost as large as the values
observed at low redshift, whereas the HDS galaxies (empty circles)
have a smaller \Dn{} index betraying a smaller luminosity-weighted
age. This distinction between HDS and non-HDS galaxies is much less
clear for the $g-r$ rest-frame color (right panel of
Fig.~\ref{figure:zdists}), which can be attributed to differences in
extinction by dust. However, over the redshift interval explored by
this paper, the GDDS sample is bluer overall than the SDSS sample.
  
%\clearpage
\begin{figure*}[!tbf]
\includegraphics[width=0.5\textwidth,height=8cm]{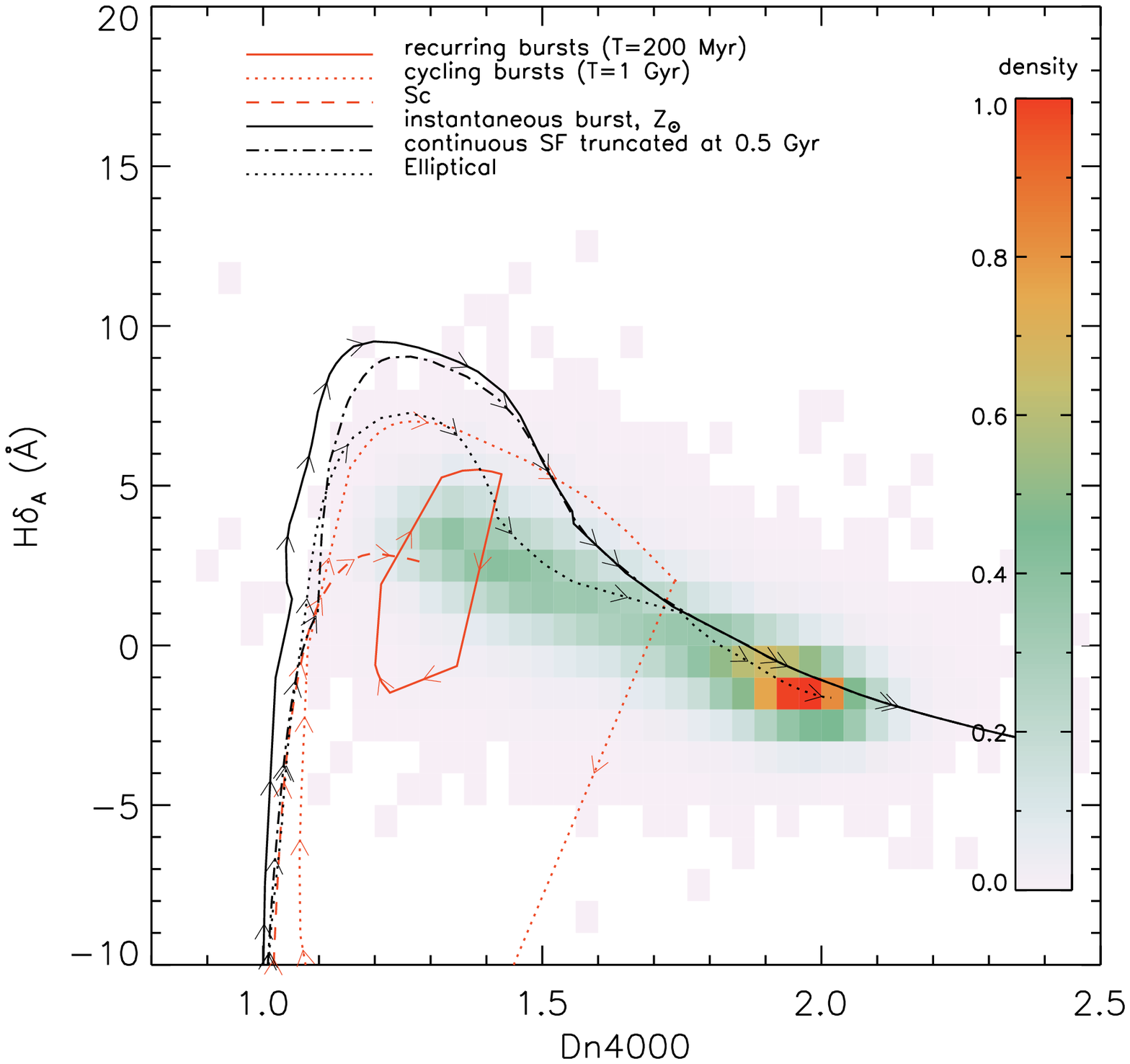} \hfill
\includegraphics[width=0.5\textwidth,height=8cm]{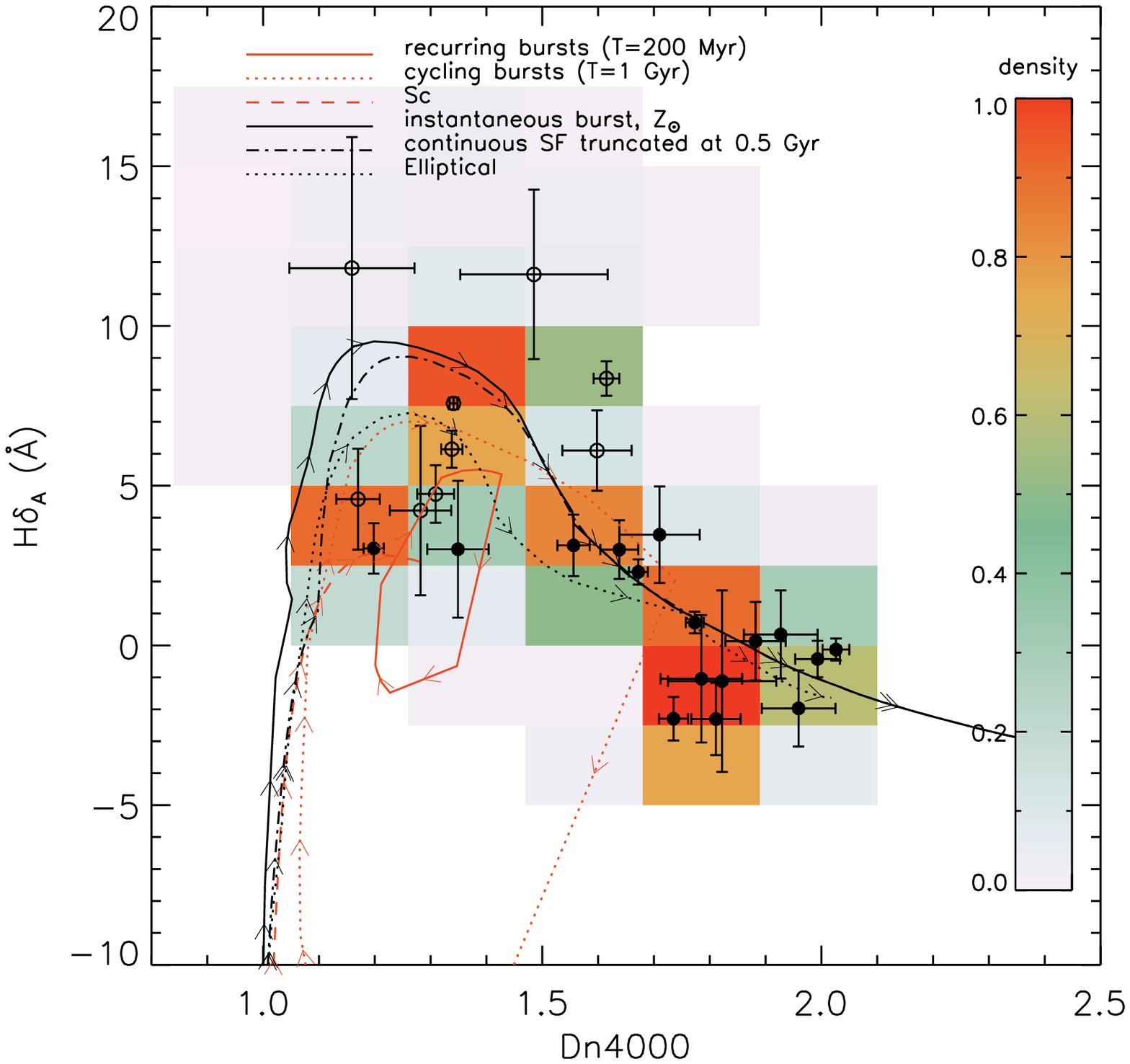}\\
\figcaption{Variation of \HdeltaA{} versus \Dn{} for SDSS galaxies (left
  panel) and GDDS galaxies (right panel).  Symbols as in
  Fig.~\ref{figure:z_Hdelta}. The arrows on the models (lines)
  indicate the direction of the evolution. For the 'cycle' models
  (closed loops), only the steady-state cycle (at $z=0$) is shown, for
  clarity.
\label{figure:Hdelta_Dn4000}}
\end{figure*}
%\clearpage

Figure~\ref{figure:Hdelta_Dn4000} shows, for each sample, the
correlation between \HdeltaA{} and \Dn{}. As will be described in the
next section, the distributions shown in this figure clearly
demonstrate that both the SDSS and GDDS samples of massive galaxies
can be divided into two categories: old evolved galaxies with a large
\Dn{} break and a small \HdeltaA{} index, and galaxies dominated by
younger stellar populations, with a smaller \Dn{} break betraying some
recent star formation activity.  The well-known bimodality in the
properties of blue and red populations of SDSS galaxies
\citep[e.g.][]{Baldryetal2004} is only faintly visible in the left
panel of this figure; bimodality is not particularly striking in our
data because our mass selection isolates the reddest part of the
bimodal distribution. Both SDSS and GDDS samples of massive galaxies
contain a large fraction of old, passively evolved galaxies (the
reddest cells in Figure~\ref{figure:Hdelta_Dn4000}).  It is once again
seen that, amongst galaxies that have recently formed stars (\Dn{}$<
1.5$), the fraction of HDS galaxies is larger in the GDDS than in the
SDSS.  Another interesting inference regarding the physical properties
of the HDS massive galaxies can be drawn from
Figure~\ref{figure:Hdelta_Dn4000}: most of the HDS galaxies seem
dominated by young stellar populations (they all have a small \Dn{}
break). 

\begin{figure*}[!tbf]
\includegraphics[width=0.5\textwidth,height=8cm]{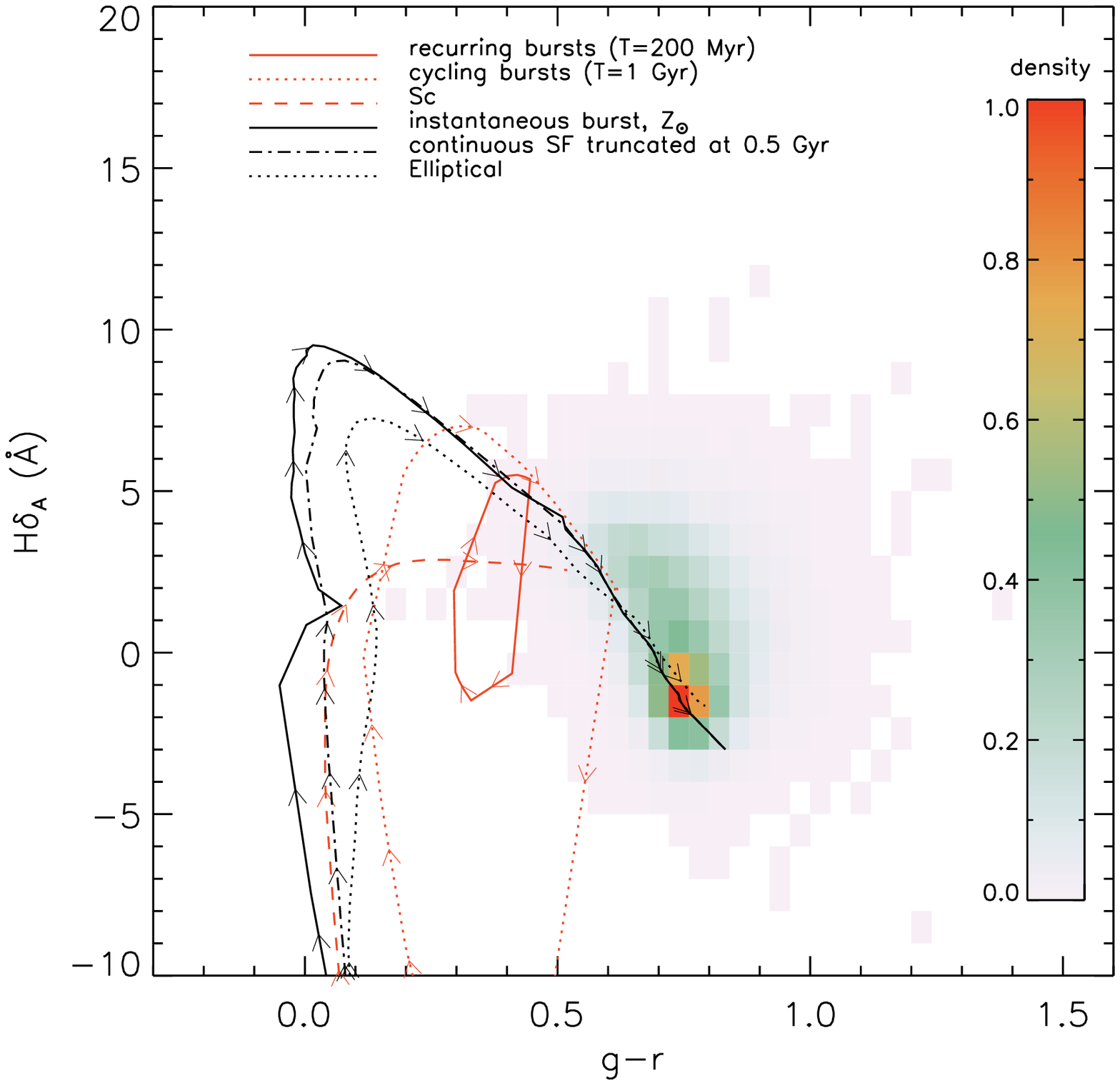} \hfill
\includegraphics[width=0.5\textwidth,height=8cm]{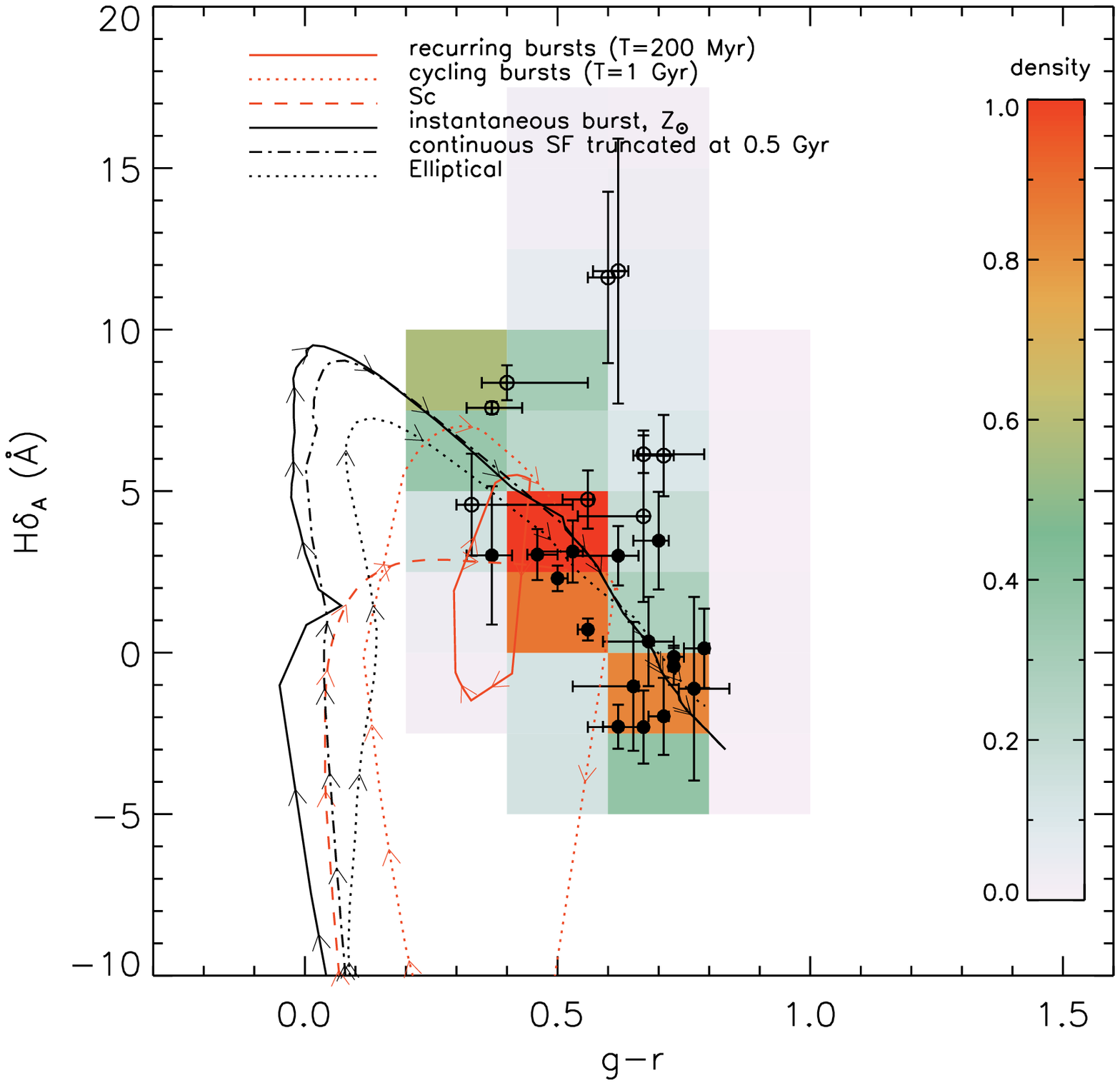}\\
\figcaption{Variation of \HdeltaA{} versus $g-r$ rest-frame color for the
  SDSS sample (left panel) and the GDDS sample (right panel). Symbols
  as in Fig.~\ref{figure:Hdelta_Dn4000}.
\label{figure:Hdelta_color}}
\end{figure*}

Figure~\ref{figure:Hdelta_color} presents correlations between
\HdeltaA{} and $g-r$ rest-frame color.  Most of the HDS objects have
redder $g-r$ colors than predicted from passively evolving models
(which include no or very little dust). This suggests that massive HDS
galaxies, from $z=1$ to the present, share a common property: they are
considerably more reddened by dust (\ebv{}=$0.2-0.5$) than evolved
galaxies. This inference is perhaps not too surprising: having
experienced recent episodes of intense star formation, they are more
likely to be dusty. This property is confirmed by a two-dimensional KS
test, implemented as in \citet{Fasanoetal1987}. The two-dimensional
statistic $D$ is 0.9, and a Monte-Carlo simulation ensures, at a
$>99.9$\% confidence level, that the data is not consistent with
dust-free models of HDS galaxies. This conclusion is in good agreement
with \citet{Shioyaetal2004} who suggest an $A_V>0.5$ mag extinction in
the red HDS systems. It is worth noting that the effect of dust on the
spectral indices  is very weak: our tests show that 
\ebv{}=0.5 would not increase \HdeltaA{} by more than 0.2~\AA{} and
\Dn{} by more than 10 percent.

%\clearpage
%\clearpage

\subsection{Evolution of the fraction of HDS systems}
Figure~\ref{figure:densities} presented evidence for a
differential evolution, from $z=1.2$ to the present, of the number and
stellar mass densities of HDS massive galaxies with respect to
populations of massive galaxies. This evolution can also be measured
and parameterized more simply through the fraction \fhds{} of HDS
galaxies in unbiased samples of massive galaxies at various redshifts.

\begin{figure*}
  \centering
  \includegraphics[width=\textwidth]{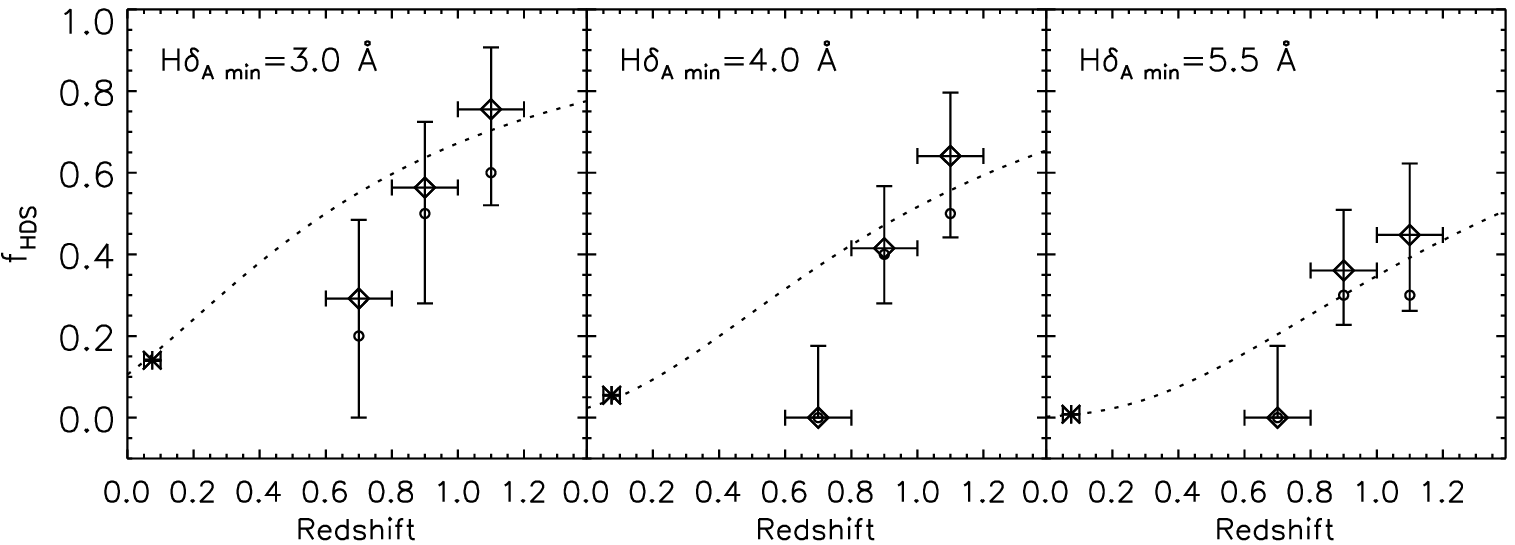}
  \figcaption{Evolution with redshift of the fraction \fhds{} of \HDS{}
  galaxies in populations of massive galaxies. Each panel corresponds
  to a different threshold for the minimum \HdeltaA{} value defining
  HDS galaxies. Triangles represent the SDSS sample.  Diamonds and
  circles represent measurements on the GDDS sample with and without
  weights, respectively. The increase of \fhds{}  with redshift (roughly, $\log_{10}(f_\mathrm{HDS}) \propto -(1+z)^{-2.5}$)
  corresponds to a $(1+z)^{2.5}$ evolution in the distribution of
  the \HdeltaA{} index. The dotted lines illustrate this evolution parameterized
  by equations~\ref{eq:fhds} and \ref{eq:alpha} (see text for details).
 \label{figure:frac}}
\end{figure*}

Figure~\ref{figure:frac} presents the evolution of this fraction and
clearly shows a steady increase of \fhds{} with look-back time, going
from \fhds{}$=0.01-0.2$ in the local universe to \fhds{}$\simeq
0.4-0.8$ at $z=1.2$.
The different panels in this figure illustrate cases where
the \HdeltaA{} threshold (used to define HDS systems) is
\HdeltaAmin{}=$3, 4,$ and $5.5$~\AA{}, and show that the measured
fraction of HDS systems increases markedly in all cases. In this
section, we propose ourselves to parameterize this evolution as a
function of the look-back time \emph{and} as a function of the threshold \HdeltaAmin{}.

%\clearpage
%\clearpage

But first, it is worth investigating how sensitive this basic result
is to the the underlying sample selection functions of the SDSS and
GDDS.  To investigate this, we explore the effect of the different
selection functions applied to the datasets. For the SDSS sample,
which is effectively complete, we can simply measure the fraction of
HDS galaxies in a single small redshift bin: $0.05<z<0.1$.  For the
GDDS sample, we define three redshift bins with $\Delta z =0.2$
between $z=0.6$ and $z=1.2$ and we measure the fraction of HDS
galaxies in each bin, weighting the contribution from each individual
galaxy by the sampling weights listed in Table~\ref{table:props}. In
effect, each galaxy is treated as if it were ``1/weight'' galaxies.
The effect of applying these weights on the measured \fhds{}, and of
ignoring them completely, can be assessed by comparing the diamonds
and circles in Figure~\ref{figure:frac}.  Clearly the effect of the
weighting by the selection function is modest. This is probably due
to the fact that the vast majority (23/25) of the massive galaxies in
our sample have an observed I-K color greater than 3 magnitudes and
lie in a region of the I vs I-K diagram where the GDDS sampling
efficiency is fairly high (cf Fig. 12 and 13 of Paper~I: about half of
the red galaxies were targeted and a redshift could be measured for
most of them).

It is worth noting that the error bars on the points shown in
Figure~\ref{figure:frac} cannot be properly described by strict
Poissonian statistics, because the quantities plotted are ratios.
Instead, we compute the uncertainties following the method of
\citet{Paterno2004} which is based on the application of the Bayes'
Theorem to binomial statistics. This method gives realistic errors
even when the total number of galaxies inside a bin is very small (and
the error on the ratios computed via Poisson errors break down), and
when the fraction is close to 0 or 1 (where strict binomial errors
break down). We choose to represent the shortest 68.3\% confidence
interval\footnote{The confidence limits derived using this method are
compatible within 4\% with the limits computed from the approximate
expressions of \citet{Gehrels1986}.}.  Another precaution is
necessary: because the error bars for the \HdeltaA{} index are quite
large for the GDDS sample, some galaxies near the limit adopted for
\HdeltaA{} could be considered as HDS or not. To quantify this effect,
we again made a Monte-Carlo simulation for each galaxy, exactly as
described in Section~\ref{Section:zdists}. For each simulation, we
measure the fraction \fhds{} and the associated lower and upper
confidence limits.  The fraction that we finally adopt is the median
value of the distribution of \fhds{}, and the final lower and upper
limits are taken as the 68\% quantiles of their respective
distributions. The final confidence interval is of the same order of
the confidence interval when only binomial statistics are considered.

In addition to showing clear evidence for evolution, Figure~\ref{figure:frac}  also
shows that the choice of the threshold for \HdeltaA{} does not
strongly influence the basic trends seen.  We will now show that it is
quite straightforward to parameterize \fhds{} as a function of both
redshift and of the threshold \HdeltaAmin{}.

%\clearpage
\begin{figure}[!tbf]
\centering \includegraphics[width=0.48\textwidth]{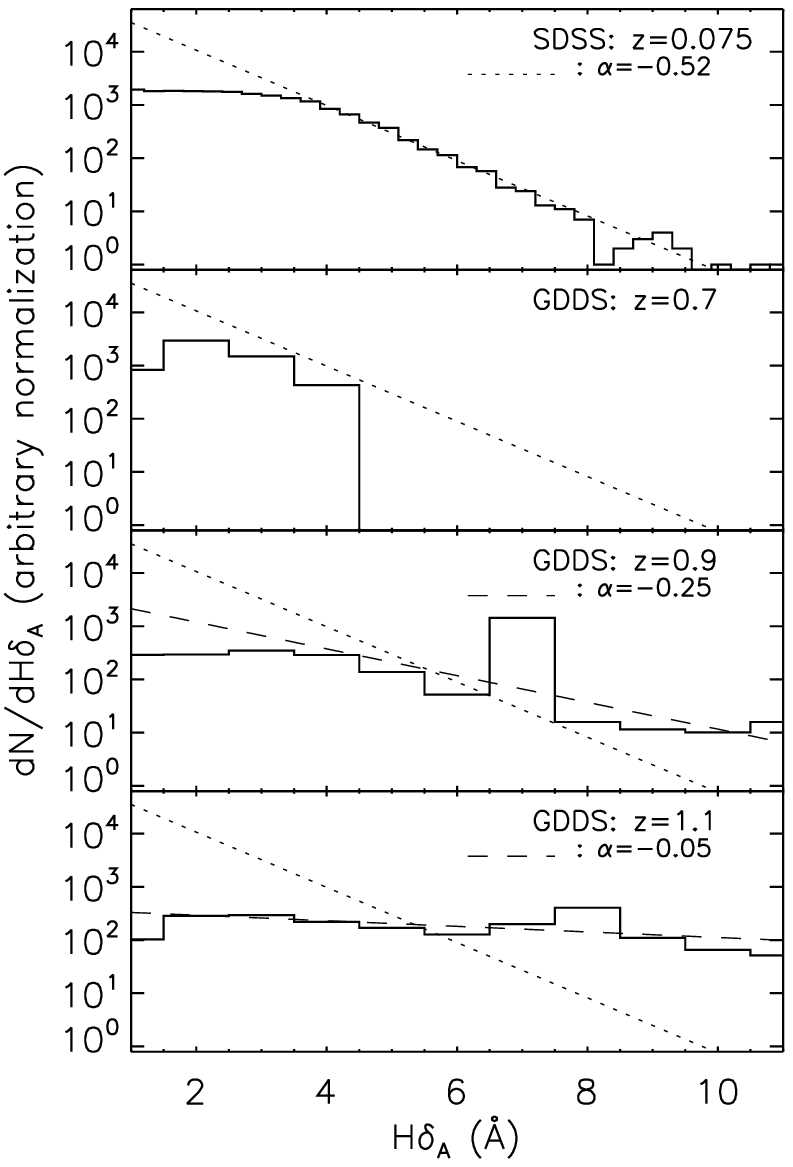}
\figcaption{Distributions of the \HdeltaA{} index measured in four
redshift bins. The slope measured on the SDSS sample (dotted line) for
\HdeltaA{}$ \gtrsim 4$~\AA{} is reported in the four panels. For GDDS
galaxies, Monte-Carlo simulations were carried out from our sample of
25 galaxies to estimate the real distribution at these redshifts; and
the dashed line represents the best-fit pure exponential
distribution. The evolution of the slope $\alpha(z)$ is shown in
Fig.~\ref{figure:slopevol}, top panel.
\label{figure:evol_number_hd}}
\end{figure}
%\clearpage

At a fixed redshift, \fhds{} obviously depends on the threshold \HdeltaAmin{}. The top
panel of Figure~\ref{figure:evol_number_hd} presents the histogram of
the distribution of \HdeltaA{} for the SDSS galaxies. It is very well
fitted by an exponential distribution (a straight line in this log-linear diagram) for \HdeltaA{}$ \gtrsim 4$~\AA{}: at the median redshift of
$z=0.075$, we have
\begin{equation}
\log_{10} \left[ \dd \mathrm{N}(z) / \dd \mathrm{H}\delta_\mathrm{A} \right] = A(z) + \alpha(z) \, \mathrm{H}\delta_\mathrm{A}
\label{eq:param}
\end{equation}
with $\alpha=-0.52$.  

We assume in the following that the form of this distribution is still valid at higher redshift. We will show that the data available at z$\simeq 1$ is compatible with this assumption.

A distribution of this form should naturally lead to a linear equation for $\log_{10} \left[ f_\mathrm{HDS}(\mathrm{H}\delta_\mathrm{A\, min}) \right]$ at a fixed redshift:
\begin{equation}
\log_{10}  \left[ f_\mathrm{HDS}(z,\mathrm{H}\delta_\mathrm{A\, min}) \right] = \alpha '(z) \left[ \mathrm{H}\delta_\mathrm{A\, min} - \mathrm{H}\delta_\mathrm{A}^0(z) \right]
\label{eq:fhds}
\end{equation}
with $\alpha ' = \alpha$ in the ideal case.  $\mathrm{H}\delta_\mathrm{A}^0(z)$ is introduced here to account for the offset of this linear equation at the redshift $z$.

\begin{figure}[!tbf]
\centering
\includegraphics[width=0.48\textwidth]{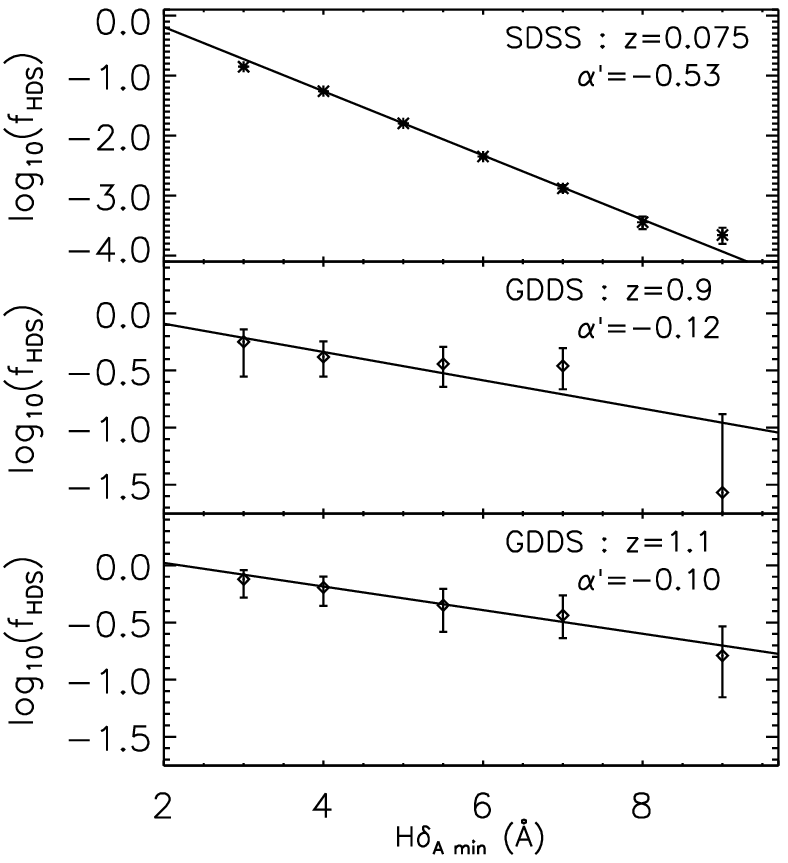}
\figcaption{Variations of \fhds{} with the threshold \HdeltaAmin{} for
three redshift bins. This figure is directly derived from
Fig~\ref{figure:evol_number_hd}, excluding the bin $0.6<z<0.8$ which does not contain enough  galaxies to allow a reliable measurement of \fhds{}. The evolution of the best-fit slope
$\alpha '(z)$  is illustrated in Fig.~\ref{figure:slopevol}, bottom
panel.
\label{figure:evol_frac_Hdmin}}
\end{figure}

The top panel of Figure~\ref{figure:evol_frac_Hdmin} shows this
quantity \fhds{} in the SDSS, as a function of the threshold
$\mathrm{H}\delta_\mathrm{A\, min}$. As expected, \fhds{} changes with
$\mathrm{H}\delta_\mathrm{A\, min}$ following equation~\ref{eq:fhds}, with
$\alpha ' =-0.53 \approx \alpha$.

%\clearpage
%\clearpage

\begin{figure}[!tbf]
\centering
\includegraphics[width=0.48\textwidth]{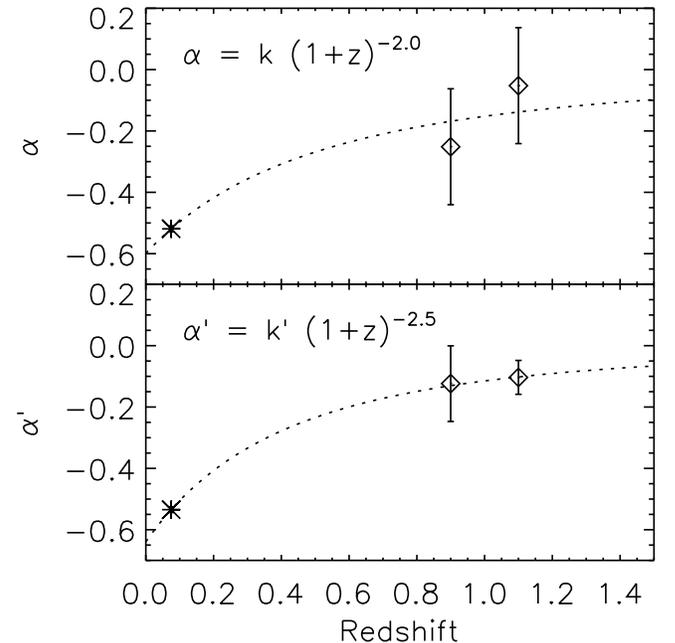}
\figcaption{Evolutions of the exponential index $\alpha(z)$ of the \HdeltaA{}
distribution (top panel) and of the exponential index $\alpha '(z)$ of
\fhds{}=f(\HdeltaAmin) (bottom panel). The values of $\alpha$ and
$\alpha '$ are reported from Fig.~\ref{figure:evol_number_hd} and~\ref{figure:evol_frac_Hdmin},
respectively. In the case of pure exponential distributions for
\HdeltaA{} (equation~\ref{eq:param}), the two panels should be identical.
\label{figure:slopevol}}
\end{figure}

At higher redshift, if the distribution of \HdeltaA{} is still a
exponential one, then the fraction of HDS galaxies should also follow equation~\ref{eq:fhds}. The three bottom panels of
Figure~\ref{figure:evol_number_hd} present the histograms of the
distributions of \HdeltaA{} for GDDS galaxies in the three redshift
bins with mean values $z=0.7, 0.9$ and $1.1$. These distributions are
the results of the Monte-Carlo simulations presented in
Section~\ref{Section:analysis}, and have an arbitrary
normalization. As for the low redshift sample, they seem to be
reasonably well described by exponential distributions, with the remarkable point that
the slope $\alpha(z)$ increases with redshift. A graphical measure of
this increase of $\alpha(z)$ is given in
Figure~\ref{figure:slopevol} (top panel).  We choose to parameterize this
evolution  by an expression of the form
\begin{equation}
\alpha(z) =k \, (1+z)^{-m}
\label{eq:alpha}
\end{equation}
and we find that $m=2.0 \pm 1.4$ and $k=-0.6$.

%\clearpage
%\clearpage

As we said before, if the distribution of \HdeltaA{} remains close to
a exponential one at high redshift, the direct measure of \fhds{} as a
function of \HdeltaAmin{} should evolve following
equation~\ref{eq:fhds}. In other words we should find that $\alpha
'(z) \approx \alpha(z)$.  The two lower panels of
Figure~\ref{figure:evol_frac_Hdmin} present our measurements of
\fhds{} as a function of \HdeltaAmin{} for the two highest redshift
bins. We observe that at the three redshifts z$\simeq$0.1, 0.9 and 1.1, \fhds{} indeed follows
Equation~\ref{eq:fhds}, with $\mathrm{H}\delta_\mathrm{A}^0 \approx
1.5$~\AA{} in every case. (An explanation for the interesting property
that this value seems to be a constant is reserved to
further investigation.)
 Comparing with the previous figure, we do indeed see that $\alpha
'(z) \approx \alpha(z)$ in these three redshift bins. Figure~\ref{figure:slopevol} (bottom panel)
shows the evolution of $\alpha '(z)$, and the best fit is obtained for
$m'=2.5 \pm 0.7$ and $k'=-0.65$, if we use the same form as in
equation~\ref{eq:alpha}.  The values of $m$ and $m'$ are compatible within their
error bars but they are slightly different. We can think of two reasons for this (small) difference. Either the
distributions of \HdeltaA{} are not really exponentials at high
redshift, or the difference is simply due to the smallness of our
sample. A larger sample would be useful to refine these values. In any
case, if the distribution is indeed a exponential, then we believe
that $m'$, derived from the fits of
$f_\mathrm{HDS}(z,\mathrm{H}\delta_\mathrm{A\, min})$, is actually a
more robust number to measure than $m$, simply because it measures a
\emph{cumulative} quantity.
 The best-fit evolution of \fhds{} (with the parameters $m'=2.5$,
$k'=-0.65$ and $\mathrm{H}\delta_\mathrm{A}^0=1.5$~\AA{}) is shown
explicitly in Figure~\ref{figure:frac} (dotted lines) and matches the
data quite well.  One might notice that the data point for the
redshift bin $z=0.6-0.8$ is slightly below the best-fitting idealized
curves: in this particular redshift bin, our sample contains only four
galaxies, and this small number might not be statistically very
significant. Moreover, the proposed parameterization should only be
valid for \HdeltaAmin{}$\gtrsim 4$~\AA{}, and the left panel of
Figure~\ref{figure:frac} is outside this range, with
\HdeltaAmin{}=3~\AA{}. Therefore, a good agreement is not expected in
this case.

To summarize, we have shown that the fraction of HDS galaxies can be
parameterized by equations~\ref{eq:fhds} and~\ref{eq:alpha}: \fhds{}
depends both on redshift and on the threshold \HdeltaAmin{}, but these
two variables are independent. In particular, we stress that the
evolution of \fhds{} with redshift does not depend on the threshold
\HdeltaAmin{} chosen to define HDS galaxies, as long as
\HdeltaAmin{}$\gtrsim 4$~\AA{}.  We have also shown that this
evolution of \fhds{} is a direct consequence of an evolution of the
\HdeltaA{} distribution. This distribution  is very well described by an
exponential distribution at low redshift, and reasonably well in the range $0.8<z<1.2$,
with an inverse slope $-1/\alpha$ increasing roughly as $~\sim
(1+z)^{2.5}$. As a result, $\log_{10}(f_\mathrm{HDS})$ rises roughly proportionally to
$-(1+z)^{-2.5}$, independently of the threshold in \HdeltaA{} that is
chosen (if \HdeltaAmin{}$\gtrsim 4$~\AA{}). However, we must add an important caveat to this result:
the parameterization by Equations~\ref{eq:fhds} and \ref{eq:alpha} is purely
empirical, and constrained only by observations at $z\simeq 0.1$ and 
$0.8\lesssim z<1.2$. Any extrapolation to earlier epochs is hazardous. In
particular, these equations suggest a never-ending increase of
\fhds{} with redshift, which is clearly not realistic  when
the universe is younger than a few $10^8$ years and galaxies  cannot be in an HDS
phase yet.

\section{Modeling and Interpretation}
\label{Section:models}
To analyze the observed colors and spectral indices shown in the
previous figures, we  use the {\sc P\'egase.2} spectral synthesis
code \citep{FRV97} for broad-band colors and the 4000\AA{} break,
together with its version at higher spectral resolution in the
optical, {\sc P\'egase-HR} \citep{PEGASEHR}, needed to measure
Lick indices. The predictions of Lick indices with {\sc P\'egase-HR}
and their accuracy with respect to other models are fully described in
\citet{PEGASEHR}. The differences between the various models available
today are small enough to reinforce the robustness of the results in
this paper.

In this section, we will compare the observed data 
with two classes of simple models: 
\begin{itemize}
\item ``Quiescent" models, defined by smooth star formation
  histories, with continuous and modest star formation rates
  throughout the life of a galaxy. 
  
 \item ``Burst'' models, which correspond to galaxies that
 build up stellar mass in a few short and intense episodes of star formation.
\end{itemize}
  
An archetypal example of a quiescent model is that described by
\citet{ZPEG} as doing a fairly good job of reproducing the optical and
near-infrared colors of local Sc spiral galaxies.  In this scenario a
galaxy is fed by infall of zero-metallicity gas with a $\tau=8$~Gyr
e-folding time-scale. Inside the galaxy, the gas is simultaneously
converted into stars following a Schmidt law, with a specific
efficiency $\nu=0.1$~Gyr$^{-1}$ ($1/\nu$ being the time required to
convert one solar mass of gas into one solar mass of stars).  A
continuous star formation history can also be approximated by a series
of small successive starbursts. The duty cycle of the bursts is
encompassed by two evolutionary scenarios, hereafter referred to as
``cycling'', or ``recurring bursts''.  In these models episodic events,
which mimic constant star formation, occur during 100~Myr. These are
followed by periods of inactivity, lasting 100~Myr in one case
(recurring bursts), and 0.9~Gyr in the other case (cycling).  It can
certainly be argued that our cycling model can hardly be considered to
be \emph{continuous} star formation, and in fact it might be better to
think of the cycling model as representing a star-formation history
that is intermediate between the quiescent mode described here, and
the ``burst'' mode described next.

In contrast to our quiescent models, our burst models are intended to
more closely reproduce the colors of local elliptical galaxies.  In
these models short time-scales ($\nu=\tau=0.1$~Gyr) are associated with
the star-formation activity in the first three Gyr. After three Gyr,
when the galaxy has converted most of the available gas into stars,
galactic winds eject all the remaining gas from the galaxy, preventing
further star formation.  Two other, more extreme, examples of burst
scenarios are a single instantaneous burst with solar metallicity, and
a period of constant star formation truncated 0.5~Gyr after the
collapse of the galaxy.
  
Throughout the remainder of this paper we will set a redshift of
$z=1.4$ as the epoch of the last major star formation episode for all
the burst-mode models considered. Clearly this redshift does not
reflect the real redshift of formation for the galaxies being modeled,
which we cannot strongly constrain on account of age-metallicity
degeneracy. Instead $z=1.4$ is simply a fiducial redshift that is
supposed to mark roughly the end of the latest major event of active
star formation that took place in the HDS massive galaxies. An
instantaneous burst model with a redshift of formation $z=2$ is also
considered and shown in Figures \ref{figure:z_Hdelta} and
\ref{figure:zdists}, to account for the old and red population of
galaxies seen at $z \simeq 1$. In all these models, nebular emission
lines are added to the stellar continuum. The fluxes in these lines
are computed with {\sc P\'egase.2} for a mean \ion{H}{2} region
consistently with the on-going and recent star-formation activity in
the galaxy. When the star-formation activity is intense, these lines
fill-in the absorption stellar lines and often lead to very negative
values for the [OII] and \Hdelta{} equivalent widths.

Tracks corresponding to a representative set of quiescent and burst
models are superposed on
Figures~\ref{figure:z_Hdelta}--\ref{figure:Hdelta_color}.  The tracks
shown on Figure~\ref{figure:z_Hdelta} are particularly illuminating,
as these suggest an evolutionary link between the $z=1$ massive HDS
galaxies and the local massive and red galaxies. Indeed, the HDS
galaxies that are already massive at $z=1$ can only be identified as
the progenitors of some local massive galaxies: they cannot lose much of their
stellar mass (which can be caused by the death of a fraction of the
stellar population) during their subsequent evolution.  
Therefore, the distant HDS galaxies necessarily must be linked to the present
 massive galaxies, most of which are passively evolving.  {\em It is clear that this
evolution can only be modeled by a break in the star-formation
history}:  the $z\approx 1$ HDS galaxies are just leaving a
burst mode with intense star formation activity (suggested by their already 
large stellar masses) which ended $\simeq 0.5-1$~Gyr
before the epoch of observation, at $z\simeq 1.2-1.5$. When seen 1-2 Gyr 
after the burst, at $z\simeq 0.8$, they are evolving almost
passively with little star formation, as described in more details
below.

We emphasize that the bulk of the massive SDSS galaxies seen at
redshift $z=0.05-0.1$ have spectral features characteristic of evolved
galaxies with very little star formation activity.  Their small
\HdeltaA{} (Fig.~\ref{figure:z_Hdelta}), large \Dn{} and red $g-r$
color (Fig.~\ref{figure:zdists}) are all well modeled by a passively
evolving stellar population, as illustrated by the models shown in
Figures~\ref{figure:z_Hdelta}--\ref{figure:Hdelta_color}. High-redshift
analogs to these objects are seen at $z\sim1$ in the population of
GDDS galaxies marked by solid circles in
figures~\ref{figure:z_Hdelta}. These objects present the same small
\HdeltaA{} values, large \Dn{} indices, and red $g-r$ colors as
evolved galaxies in the local universe\footnote{Note that the stellar
mass cutoff we chose when defining our sample corresponds (roughly) to
the pivot mass in the observed bimodality between blue and red
galaxies in the local universe
\citep{Baldryetal2004,Kauffmannetal2003}.}. These objects connect
directly onto a number of recent studies of massive passively evolving
galaxies out to even higher redshifts
\citep[e.g.][]{GDDS_PIV,GDDS_PIII,Cimattietal2004}.   

 A comparison of the ``quiescent'' models (which all have a smoother
star-formation history than the burst models) with the data is also 
interesting. Figure~\ref{figure:z_Hdelta} shows that continuous
star-formation for a long period of time (Sc model) leads to
intermediate values of the \HdeltaA{} index ($2-3$~\AA). This is not
true if the star-formation is intermittent: when the star-formation
episodes are spaced by more than 100 Myr (which is the case of our
recurring and cycle models), the \HdeltaA{} index shows substantial
variations from very positive (if star-formation is halted abruptly)
to very negative values (because of nebular emission lines). The cycle
and recurring burst models shown in Figure~\ref{figure:zdists} also
make it clear that if the star formation episodes are too frequent
(i.e. repeated on periods smaller than a very few Gyrs), the large
\Dn{} indices and red rest-frame colors of the massive galaxies at
$z=0$ or $z=1$ can never be reached.

A particularly nice aspect of modeling \HdeltaA{} and \Dn{} is that
models used for interpretation of these two quantities are
sufficiently robust that uncertainties in them are unlikely to lead to
wild misinterpretations of the data.  Indeed, we have checked that a
change in the IMF, within reasonable limits, does not affect the
predicted \HdeltaA{} index by more than 2~\AA{} and \Dn{} by more than
0.2. To some extent, the uncertainties in the stellar evolution tracks
can be explored by looking at the differences between the prediction
of the {\sc P\'egase} models and the Galaxev models
\citep{BC2003}. \citet{PEGASEHR} showed that the differences in the
predicted \HdeltaA{} indices are smaller than 1~\AA{} at any
metallicity. We also verified that the differences on the predicted
\Dn{} index are smaller than the errors in the data. However,
uncertainties on the effect of the non-solar abundances of $\alpha$
elements remain unclear and might be responsible for slight deviations
of the model tracks from the data. The general trends will, however,
remain robust.

It is worth emphasizing that strong \Hdelta{} in absorption is,
strictly speaking, a signature of a recent {\em break} in
star-formation, rather than of any preceding star-burst activity. 
Moreover, a complete halt in the star-formation activity is not needed
to obtain an HDS system. It is the contrast between the current and recent-past 
star-formation rates that must be large.

%It {\bf would be misleading} to refer to HDS systems as `post-starburst'
%galaxies, since ordinary late-type spirals would exhibit this
%signature if their star-formation was simply halted. Referring to them
%as `post-star-formation' galaxies would be more accurate, although the
%`post-starburst' taxonomy is rather entrenched.  
%A more physical way
%to look at this issue is

To illustrate the importance of this contrast, we can show that the current star-formation rate is small in comparison to
its average past value by comparing the
characteristic growth time-scale of a galaxy to the age of the
universe at its redshift. Following Paper~V, we define this growth
time-scale (expressed in Gyr) as the ratio $M_\ast / SFR$: it is the
time needed for a galaxy to assemble its current mass if its SFR is
fixed at its current value. If the growth time-scale of a galaxy is
larger than the age of the universe at its redshift, we can conclude
that such a galaxy is in a mode of passive evolution (relative to its
past star-formation history).  The SFR values of our high redshift
galaxies, estimated from the [OII]$\lambda 3727$ emission line and
assuming A$_V$=1, are listed in Table~\ref{table:props}. The typical
values of the on-going SFR are 1-2~$M_\odot$\,yr$^{-1}$ for HDS
galaxies (and $0.1-1$~$M_\odot$\,yr$^{-1}$ for the non-HDS, ``older''
galaxies). Although these values are non negligible, they are still
small with respect to the past mean SFR: Table~\ref{table:props} shows
that $M_\ast / SFR$ is larger than the age of the universe, in a ratio
of $\simeq 3$ for the HDS galaxies, and $\simeq 10$ for the non-HDS
galaxies. This is in good agreement with the results of Paper V, and
it means that the massive galaxies are already in a mode of passive
evolution at $z=1$. It also illustrates the large contrast
between the recent ($\simeq 1$ Gyr ago) large bursts of star formation
and the modest values of the on-going SFR in HDS systems.

%\clearpage
\begin{figure}%[!tbf]
  \centering
  \plotone{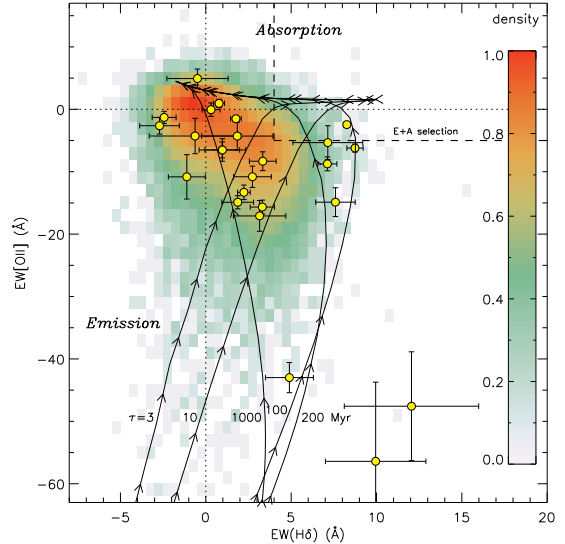} %\hfill
  \figcaption{Selection of E+A galaxies (defined by the dashed box:
    \HdeltaEW{}$>4$~\AA{} and \OIIEW{}$>-5$~\AA{}). The shaded colored
    region corresponds to the SDSS sample (same color-coding as in
    Fig.~\ref{figure:z_Hdelta}). The individual GDDS galaxies of our
    sample are shown on the top of the SDSS sample, with their error
    bars. The evolutions predicted by closed-box exponential models
    (solid lines, with arrows marking every doubling in age) are given
    for various e-folding time-scales $\tau$. The fraction of galaxies
    in the E+A phase (or simply in the HDS phase with
    \HdeltaEW{}$>4$~\AA{}) is much larger in the GDDS sample than at
    low redshift. Moreover, only models corresponding to a rapid decrease of the star-formation rate ($\tau \lesssim 300$~Myr) reach the
    E+A phase (see also Table~\ref{table:HdeltaEW_ages}).
  \label{figure:HdOII}}
\end{figure}
%\clearpage

The classical way to study ``post-starburst" galaxies (i.e. with
completely truncated star formation) is to focus on the sub-sample of
HDS galaxies with little (or no) emission lines, as measured by their
equivalent widths. The most well-known categorization of these
galaxies is the `E+A' population
\citep{Dressler&Gunn1983,Baloghetal1999,Poggiantietal1999,Zabludoffetal1996},
with strong \Hdelta~ absorption lines coupled with little (or no)
[OII] emission.  To facilitate comparison with previous work, we
conclude this section by investigating the fraction of `E+A' galaxies
in our HDS population.  Table~\ref{table:props} includes measurements
of [OII] equivalent width, with emission lines represented by negative
values\footnote{Note that, for this analysis, we chose to use the
definition of the equivalent width \HdeltaEW{} given in
Table~\ref{table:indices} instead of the Lick \HdeltaA{} index used
earlier in this paper in order to make our results directly comparable
to previous work \citep{Abrahametal1996,Baloghetal1999,
Gotoetal2003,Tranetal2004}.  When measured on the SDSS sample, the
average value of \HdeltaEW{}$-$\HdeltaA{} is $0.50$~\AA, with an rms
dispersion of $0.78$~\AA{}.}. As described below, our GDDS sample
contains one or two E+A galaxies, depending on how error bars are treated.

Figure~\ref{figure:HdOII} shows the distribution of the two samples of
massive galaxies on a \HdeltaEW{}-\OIIEW{} diagram. The SDSS sample is
represented with the same color-coding of the number density as in
previous figures. The individual GDDS galaxies are shown with their
error bars. Some very simple closed-box evolutionary models computed
with {\sc P\'egase} are over-plotted: these models have exponentially
decreasing star-formation rates with e-folding time-scales $\tau=3, 10,
100, 200$ and $1000$~Myr (Column 1 of
Table~\ref{table:HdeltaEW_ages}). All these models form stars at solar
metallicity. The arrows give the direction of the evolution in the
diagram: \OIIEW{} is highly negative soon after the maximum of the
star formation because of the presence of strong emission lines, and
increases continuously afterward.  The models with $\tau <300$~Myr
pass through an HDS phase (if defined by \HdeltaEW{}$>4$~\AA) and an
E+A phase (if defined by an HDS galaxy with \OIIEW{}$>-5$~\AA{}) which
is illustrated by the dashed line in Figure~\ref{figure:HdOII}.

\begin{table}
\caption{Closed-box models with exponentially decreasing SFR.}
\label{table:HdeltaEW_ages}
\tablewidth{0pt}
\centering
\begin{tabular}{lcccc}
\hline
\hline
$\tau$ &   \HdeltaEW{}  & Age &  $\Delta t_\mathrm{HDS}$ & $\Delta t_\mathrm{E+A}$\\
(Gyr)  &     (\AA)      & (Gyr) &      (Gyr)             &   (Gyr) \\
\hline
0.003       & 10.0   &   0.25 & 1.56 & 1.56 \\
0.01      & 10.0   &   0.30 & 1.54 & 1.53 \\
0.1     & 8.8   &   0.60 & 1.48 & 1.11 \\
0.2     & 7.1   &   0.80 & 1.51 & 0.67 \\
1.0    & 3.5   &   1.80 & 0.00 & 0.00 \\
\hline
\end{tabular}
\end{table}

The maximum \HdeltaEW{} values reached by our models, and the
corresponding ages, are given in Table~\ref{table:HdeltaEW_ages}, Columns 2 and 3. This
table also records (Columns 4 and 5) the total time spent in the HDS and E+A phases.
The table shows that  (i) the maximum \HdeltaEW{} is reached
sooner for rapidly decreasing SFRs, (ii) the time spent in the HDS
phase is almost independent of the e-folding time-scale when this phase
is reached, and (iii) a given galaxy spends less time in the E+A phase
when the SFR declines more slowly. On this basis, it is clear from
Figure~\ref{figure:HdOII} that the proportion of galaxies showing signs
of recently truncated star formation ($\tau < 200$ Myr and
\HdeltaEW{}$>4$~\AA) is higher for the GDDS sample than for the SDSS
sample.

Models in Table~\ref{table:HdeltaEW_ages} and
Figure~\ref{figure:HdOII} can be used to make a theoretical prediction
regarding the evolution of E+A systems: while models with $\tau<
300$~Myr have both an HDS and an E+A phase, all models that have an
HDS phase do not necessarily have a E+A phase. (For instance, models
with $0.3< \tau < 1.0$~Gyr). Therefore, depending on how fast the
star-formation is halted, one might expect that the evolution of the
fraction of E+A galaxies would show milder evolution than the fraction
of HDS galaxies.  Strictly speaking, $4_{-3}^{+5}$\% (1/25) of
the GDDS galaxies are E+A at $z=0.6-1.2$ and 3.1\% (1561/50,255) of
the SDSS sample at $z=0.05-0.1$ meet the E+A definition. (The GDDS
galaxy 02-1543 might be counted as an additional E+A, depending on
how one treats its error bar.) The fractions at low and
high redshifts being roughly compatible,  we don't observe
any evolution in the fraction of E+A galaxies. However, the
small number statistics involved and our large error bars leave
room for further studies on this topic.

%Although the fractions at low and high
%redshift seem roughly compatible, the small number
%statistics involved make it impossible to conclude anything about a hypothetical 
%evolution in the fraction of E+A galaxies.

It is of interest to note that a number of recent papers derive a
$(1+z)^m$ evolution to describe the rise in the merger rate as a
function of redshift, with $m$ indices that are similar to the
exponential index we obtain for the rise in the fraction of HDS
systems as a function of redshift: compare our value of $m = 2.5 \pm
0.7$ with $m=3.4 \pm 0.6$ obtained by \cite{Lefevreetal2000} and
$m=2.5 \pm 0.7$ obtained by \cite{Pattonetal2002}.  However, the
evolution in the merger rate is poorly constrained.  For example,
larger values are also obtained, such as $m=4 \pm 1$ by
\citet{Reshetnikov2000} and $m=4-6$ according to
\citet{Conseliceetal2003}.  At this stage we think the most that can
be said is that our results may hint at a relationship between merging
and the HDS phase. A similar relationship is suggested by
the recent work of \citet{Gotoetal2005} on 266 E+A galaxies in the
SDSS.  If a link exists between HDS galaxies and mergers, the
morphologies of HDS galaxies should reflect these dramatic
events.  Figure~\ref{figure:allspectra} shows that many of the HDS
galaxies (left column) indeed have disturbed morphologies which can be
signs of merging. Moreover, the ratio between irregular and relaxed
morphologies should roughly scale as the ratio of time-scales for
merging ($\simeq 0.5$~Gyr) and spectral evolution ($\simeq
0.3-2$~Gyr). This statement seems generally consistent with
Figure~\ref{figure:allspectra}, which shows several HDS galaxies with
almost relaxed morphologies (02-1724, 02-1777, 22-0315), but further
discussion of this topic is reserved for an upcoming paper in this
series (Abraham et al. 2006, in preparation).

%----------------------------------------
\section{Conclusions}
%----------------------------------------
\label{Section:conclusion}
By comparing two mass-selected samples of galaxies at low (SDSS) and
high (GDDS) redshift, we find that the fraction of galaxies with
strong H$\delta$ absorption lines increases with redshift up to
$z=1.2$. This evolution can be parameterized as a function of redshift:
$(1+z)^{2.5 \pm 0.7}$. It does not depend on the threshold
chosen for \HdeltaA{} to define HDS galaxies, and it reflects a similar
evolution for the distributions of the \Hdelta{} equivalent widths. By
modeling this variation using {\sc P\'egase} models as a function of
\Dn{} and the rest-frame $g-r$ color, we conclude that this evolution
originates from a fundamental change in the mode of star formation in
massive galaxies that occurred somewhere around $z\simeq 1.4-2$.  Our
models suggest that the star formation rates drop sharply in most
massive galaxies at that epoch, which is consistent with the measures
of star formation rate densities presented in Paper V. 

This drop in the star formation rate, together with the already large
stellar masses of these $z\simeq 1$ galaxies, suggests very active
phases in the star-formation at higher redshifts.  The massive
obscured starbursts being detected with the Spitzer Space Telescope at
$z \simeq 2-2.5$ could very well be the progenitors of our sample of
$z\simeq 1$ galaxies: the ``old'' galaxies seen in our sample at
$z=1$, with small \HdeltaA{} indices and large \Dn{} breaks, have
necessarily experienced their last significant episodes of
star-formation at $z \gtrsim 1.8$ and assembled most of their stellar
mass before that time. These episodes must have been very intense (at
least $\simeq 30$~$M_\odot\, yr^{-1}$ on average), given the large
stellar mass of these galaxies and the age of the universe at this
redshift.  Moreover, some of the $z=2-2.5$ starbursts could also have
spanned larger timescales ($2-3$~Gyr) in which case they could become
HDS galaxies at $z\simeq 1$, on the condition that their
star-formation declined at $z \simeq 1.4$, on time-scales shorter than
a few hundred Myrs. This picture is in fairly good agreement with
the recent findings of \citet{Treuetal2005} who estimate
that the massive field spheroidal galaxies seen at $z=0.2-1.2$ in the
GOODS-N field formed most of their stellar mass at $z\gtrsim 2$, with
subsequent activity continuing to lower redshifts.

Finally, our study of spectral features in massive galaxies from
$z=1.2$ to the present shows that the redshift range $z=1.4-2$
corresponds to the epoch at which the star-formation in these galaxies
has changed from an active phase to a passive one.  An echo of this
change in the mode of star-formation activity is imprinted in the
strong H$\delta$ absorption lines seen in massive galaxies at $z\sim
1$.

\section{Acknowledgments}
The authors wish to thank Jarle Brinchmann for many stimulating 
discussions and for sharing with us a number of plots which motivated
much of this work. We are also in debt to the anonymous referee whose comments helped to improve the quality of this paper. Observations were obtained at the Gemini Observatory,
which is operated by AURA Inc., under a cooperative agreement with the
NSF on behalf of the Gemini partnership: the NSF (US), PPARC (UK), NRC
(Canada), CONICYT (Chile), ARC (Australia), CNPq (Brazil) and CONICET
(Argentina) and at the Las Campanas Observatory of the OCIW.  
RGA acknowledges support from NSERC and the Province of Ontario.
KG and SS acknowledge support from the David and Lucille Packard Foundation.
    Funding for the creation and distribution of the SDSS Archive has been provided by the Alfred P. Sloan Foundation, the Participating Institutions, the National Aeronautics and Space Administration, the National Science Foundation, the U.S. Department of Energy, the Japanese Monbukagakusho, and the Max Planck Society. The SDSS Web site is http://www.sdss.org/.
    The SDSS is managed by the Astrophysical Research Consortium (ARC) for the Participating Institutions. The Participating Institutions are The University of Chicago, Fermilab, the Institute for Advanced Study, the Japan Participation Group, The Johns Hopkins University, the Korean Scientist Group, Los Alamos National Laboratory, the Max-Planck-Institute for Astronomy (MPIA), the Max-Planck-Institute for Astrophysics (MPA), New Mexico State University, University of Pittsburgh, University of Portsmouth, Princeton University, the United States Naval Observatory, and the University of Washington.

%bibliography
\bibliographystyle{aa}

\end{document}